\documentclass[a4paper]{article}
\usepackage[utf8]{inputenc}
\usepackage[T1]{fontenc}
\usepackage{mathptmx}
\usepackage{graphicx}
\usepackage{amsmath,amssymb,amsthm,bm}
\usepackage{natbib}
\usepackage{dcolumn}
\usepackage{multirow}
\usepackage{subcaption}
\usepackage{comment}
\usepackage{etoolbox}
\usepackage[mathscr]{eucal}
\usepackage{xcolor}
\usepackage{todonotes}
\usepackage{booktabs}
\usepackage[margin=1.2in]{geometry}

\title{Parametric Reduced-Order modeling and Closed-Loop Control of Tandem-Cylinder Wakes}
\author{Tea Vojkovi\'c\textsuperscript{1,2}, Dimitris Boskos\textsuperscript{2}, Abel-John Buchner\textsuperscript{1}\\[0.5em]
\small \textsuperscript{1}Laboratory for Aero and Hydrodynamics, Delft University of Technology, The Netherlands\\
\small \textsuperscript{2}Delft Center for Systems and Control, Delft University of Technology, The Netherlands\\[0.5em]
\small Corresponding author: t.vojkovic@tudelft.nl}
\date{April 2, 2026}
\begin{document}
\maketitle
\begin{abstract}
The flow around two circular cylinders arranged in a tandem exhibits complex wake interactions that lead to amplified unsteady loads, particularly in the co-shedding regime where a fully developed wake forms in the gap between the cylinders. Although various control strategies have been proposed to mitigate these effects, most prior studies have focused primarily on load alleviation. Complete suppression of vortex shedding, both in the gap region and in the wake of the second cylinder, has so far only been achieved using open-loop approaches. In this work, we propose a closed-loop control framework for suppressing vortex shedding in tandem-cylinder flows in the co-shedding regime. Focusing on low Reynolds numbers and sufficiently large spacings, we derive a parametric reduced-order model using a global weakly nonlinear analysis of the incompressible Navier–Stokes equations. The model is generalized to account for time-dependent forcing and facilitates the real-time prediction of the flow evolution. Using this model, we design a model predictive controller and apply it to the full-order system via velocity measurements and volumetric forcing. The approach is demonstrated for a cylinder spacing of eight diameters. Vortex shedding is fully suppressed in both the gap region and the downstream wake for Reynolds numbers $Re = 50, 60$, and $70$, while a significant reduction in flow unsteadiness is achieved at $Re = 80$. We further show that effective control is possible with limited sensing: suppression is achieved using a single measurement point for $Re = 50$ and two-point measurements for $Re = 60$ and $70$.

\end{abstract}

\section{Introduction}

Multiple cylindrical structures immersed in a flow are common in engineering applications, ranging from large-scale systems such as oil pipelines, offshore platform columns, and bridge piers to smaller-scale devices including heat exchangers. Depending on the application, these structures may be arranged in side-by-side, staggered, or tandem configurations.

The flow around two equal-diameter circular cylinders arranged in tandem is a benchmark problem for studying wake interaction in multi-body flows. Owing to its practical relevance and rich flow physics, this configuration has been widely investigated experimentally and numerically. A comprehensive review of these studies is provided by~\citet{sumner2010two}.

Based on experimental studies~\citet{zdravkovich1987effects} and later~\citet{xu2004strouhal} and~\citet{zhou2006flow} classified the flow around tandem cylinders into three regimes according to the spacing ratio $\gamma = L/D$. Here, $L$ denotes the centre-to-centre spacing between the cylinders and $D$ their diameter. The reported regime boundaries differ primarily depending on the Reynolds-number ranges considered.

The following regimes are identified:
(i) \textit{The extended-body regime} occurs at small spacings, typically $1 < \gamma < 1.2-1.8$ in the experiments of~\citet{zdravkovich1987effects}, or $1< \gamma < 2$ in the study of~\citet{zhou2006flow}, and is characterized by the two cylinders behaving as a single bluff body; the shear layers separated from the upstream cylinder enclose the downstream cylinder without reattachment and roll up to form a K\'arm\'an vortex street behind it, while the gap flow remains largely stagnant \cite{zdravkovich1985flow}, with occasional cavity-type oscillations \cite{hetz1991vortex}. 
(ii) \textit{The reattachment regime} is observed at intermediate spacings, approximately $1.2-1.8 < \gamma < 3.4-3.8$ in~\citet{zdravkovich1987effects} or $2 < \gamma < 5$ in~\citet{zhou2006flow}, where the increased separation allows the upstream shear layers to reattach intermittently onto the downstream cylinder, generating unsteady gap vortices; within this regime,~\citet{zhou2006flow} further noted that for $2 < \gamma < 3$ reattachment occurs predominantly on the downstream side of the second cylinder, whereas for $3 < \gamma < 5$ it occurs more frequently on the upstream side. 
(iii) \textit{The co-shedding regime} occurs at larger spacings, reported as $\gamma > 3.4-3.8$ in~\citet{zdravkovich1987effects} and $\gamma > 5$ in~\cite{zhou2006flow}, where the downstream cylinder lies outside the vortex formation region of the upstream cylinder, permitting the development of a K\'arm\'an vortex street in the gap. In this regime, both cylinders shed vortices at the same frequency~\cite{sumner2010two}. A qualitatively similar three-pattern classification has been reported in~\citet{carmo2010possible} at low Reynolds numbers, consisting of: 
(i) \textit{symmetric vortices in the gap} at small spacings (observed at $\gamma=1.5$); 
(ii) \textit{alternating gap vortices} at intermediate spacings ($2 < \gamma <3.7$--$4.6$, depending on Reynolds number); and 
(iii) \textit{fully developed wake in the gap} at larger spacings ($\gamma > 3.2$--$4$).

In the co-shedding regime, vortices which have shed from the upstream cylinder periodically impinge on the downstream cylinder, leading to larger fluctuating fluid-dynamic forces compared to those experienced by an isolated cylinder~\cite{carmo2010possible,liu2024primary}. In this work, the force perpendicular to the direction of free stream (transverse force) is referred to as lift, while the force parallel to it (streamwise force) is referred to as drag. Depending on the application context, such amplified unsteady loads may result in reduced operational efficiency, flow-induced vibration and fatigue, or increased maintenance costs. Consequently, the suppression or mitigation of vortex shedding in tandem-cylinder configurations is of considerable practical importance.

In general, the suppression or mitigation of flow instabilities such as vortex shedding may be achieved using either passive or active flow-control techniques. Passive control strategies typically rely on geometric modifications. The flow over tandem cylinders at $\gamma = 3.7$ and $Re = 1.66 \times 10^{5}$ was effectively controlled, for example, using porous coating in the combined numerical and experimental study of~\citet{liu2015tandem}. Vortex shedding in the gap was suppressed, while unsteadiness in the wake of the downstream cylinder persisted but was mitigated. In that study, achieving this performance required a parametric investigation: the coating thickness and porosity were treated as design parameters and systematically varied to identify configurations that provide effective and robust suppression.

Active flow control introduces external forcing into the flow and may be implemented using either open-loop or closed-loop strategies. Active control approaches can be based on forced dynamical models of the flow, in which the control input is determined from model-based predictions of the flow response to actuation, or on model-free formulations. Model-free approaches typically rely on trial-and-error tuning or extensive parametric studies, whereas suitable forced dynamical models can provide insight into the flow response to actuation and reduce reliance on exhaustive parameter searches. Active control of tandem-cylinder wakes has been implemented using plasma actuators, synthetic jets, and prescribed kinematics of the cylinders.

Open-loop control of tandem-cylinder wakes has been investigated using plasma actuators in experimental studies by~\citet{kozlov2011plasma, latrobe2024flow} and in numerical work by~\citet{eltaweel2014numerical}. \citet{kozlov2011plasma} and \citet{eltaweel2014numerical} considered $\gamma = 4$ at $Re = 22{,}000\!-\!172{,}000$, while \citet{latrobe2024flow} examined $\gamma = 3, 4,$ and $5$ at $Re = 4700$. In all cases, plasma actuators were applied to the upstream cylinder, 
leading to attenuation of vortex shedding from the upstream body. This, in turn, resulted in a more benign wake interaction with the downstream cylinder and a reduction in the magnitude of fluctuating fluid forces acting on it. However, vortex shedding was not fully suppressed, particularly in the wake of the downstream cylinder; rather, the controlled tandem configuration behaved similarly to that of an isolated cylinder. The actuation parameters in these studies were selected based on experimental observations and parametric exploration, rather than being derived from a forced dynamical model of the flow.
At lower Reynolds numbers,~\citet{liu2024primary} numerically employed a constant-velocity jet for open-loop control at $Re = 75\!-\!200$, considering gap spacings $\gamma = 4$ and $6$. In this study, vortex shedding in the gap region was suppressed and shedding in the wake of the downstream cylinder was strongly attenuated. The control strategy was not based on a forced dynamical model of the flow; instead, multiple jet velocities were tested to identify effective operating conditions. Furthermore, this open-loop strategy relied on constant actuation inputs, and sustaining such suppression may therefore incur high energy costs or lack robustness to changes in flow conditions. 

Closed-loop control, in contrast, relies on flow measurements to determine the control action, making it potentially more energy efficient and capable of adapting to changing flow conditions. A model-free closed-loop control strategy was experimentally applied by \citet{wolfe2003feedback} to the flow around tandem cylinders at a spacing ratio $\gamma = 4.5$, where synthetic-jet actuation was employed on the upstream cylinder and the feedback was relying on lift force measurements from the downstream cylinder. This approach achieved a significant reduction in the fluctuating load acting on the downstream cylinder.

Achieving real-time predictive closed-loop control generally requires reduced-order models that enable the rapid simulation of the flow evolution. In this context, a deep-learning-based approach was proposed by~\citet{xie2023applying} for a wall-confined tandem-cylinder configuration at Reynolds number $Re = 100$ and spacing $\gamma = 4.5$. In this numerical study, rotation of the downstream cylinder was used as the actuation mechanism, while the state was measured through an array of $45$ velocity sensors distributed in the flow field. Using this framework, up to a $75\%$ reduction in the lift force fluctuations on the downstream cylinder was achieved. However, this control strategy primarily alleviated unsteady dynamics in the gap region and did not suppress vortex shedding in the wake downstream of the second cylinder.

To the best of our knowledge, existing studies have either alleviated unsteady loading without suppressing vortex shedding in the wake of the downstream cylinder, or nearly  achieved such a suppression only through open-loop control, which typically lacks robustness guarantees. The present work therefore focuses on \emph{model-based closed-loop control} aimed at suppressing vortex shedding in the co-shedding regime, both in the gap region and in the wake downstream of the second cylinder.

To this end, we construct a low-dimensional model that enables fast prediction of the flow evolution over tandem cylinders and is suitable for real-time closed-loop control. Our analysis is restricted to Reynolds numbers for which the flow remains two-dimensional, i.e. below the onset of three-dimensional instabilities. Furthermore, we focus on sufficiently large spacings for which the primary two-dimensional instability of the tandem-cylinder wake arises through a supercritical Hopf bifurcation ($\gamma>5.5$)~\cite{mizushima2005instability}, while the downstream cylinder is still subjected to significantly higher unsteady loading than an isolated cylinder. As this bifurcation scenario is analogous to that of an isolated cylinder, we employ a global weakly nonlinear analysis of the incompressible Navier--Stokes equations following the framework introduced by \citet{sipp2007global}.
To render the resulting model applicable to closed-loop control, we generalize the formulation originally developed in \citet{sipp2012open} for open-loop harmonic forcing with constant amplitude to the case of slowly time varying forcing amplitudes.

This framework yields a two-dimensional reduced-order model that (i) captures the nonlinear forced dynamics, (ii) accounts for time-dependent forcing, (iii) explicitly depends on the Reynolds number and is valid over a finite parameter range, and 
(iv) provides mappings between the full- and reduced-order state and input spaces, enabling the implementation of feedback control in the full-order system.
In addition, (v) the model is computationally efficient, as its coefficients are obtained from the solution of a set of linear equations, avoiding the need for long time-resolved simulations typically required by data-driven approaches, and (vi) it is physically interpretable, providing modal decompositions and insight into the underlying flow dynamics.

Based on this reduced-order model, we design a model predictive controller (MPC) that is consistently mapped to a stabilizing controller for the full-order system, which is considered here in a purely numerical setting. The controller is formulated as an output-feedback law using pointwise velocity measurements, and actuation is introduced in the form of volumetric forcing. Following the MPC paradigm, an optimization problem is solved recursively to determine a sequence of forcing amplitudes over a finite prediction horizon, where the future flow evolution is predicted using the reduced-order model as a surrogate for the full dynamics.

Although the present study focuses on tandem-cylinder flows, the proposed methodology is applicable more generally to incompressible flows that lose stability through a supercritical Hopf bifurcation. We demonstrate the approach for tandem cylinders with spacing ratio $\gamma=8$ at Reynolds numbers $Re = 50, 60, 70,$ and $80$. Successful suppression of wake oscillations, both in the gap and behind the downstream cylinder, is achieved for $Re=50,60,$ and $70$ using localized forcing, with single-point velocity measurements for $Re=50$ and two-point measurements for $Re=60$ and $70$. For $Re=80$, a significant alleviation of the unsteady dynamics is obtained using full-field velocity measurements.

The paper is organized as follows. First, we introduce our problem and set the objectives of the paper in Section~\ref{sec:2}. Section~\ref{sec:WNA} addresses our generalization of the weakly nonlinear analysis and derivation of the parametric reduced-order model. 
In Section~\ref{sec:control}, we use this model to design a closed-loop controller for the flow. Section~\ref{sec:results} presents the numerical setup and demonstrates the proposed modelling and closed-loop control approach for the flow around two tandem cylinders at $Re=50,60,70,80$.

\section{Flow over two cylinders in tandem configuration}\label{sec:2}
\subsection{Governing Equations}

We consider a two-dimensional incompressible flow with freestream velocity $u_{\infty}$ past two circular cylinders of equal diameter $D$, arranged in tandem. The center-to-center spacing between the cylinders is denoted by $L$, and the corresponding spacing ratio is defined as $\gamma = L/D$. The flow is assumed to be subjected to a time-varying volumetric forcing term $\mathbf{f}(t)$.

The flow is governed by the incompressible Navier--Stokes equations, written here in non-dimensional form as
 \begin{subequations}\label{eq:NS}
 \begin{eqnarray} 
        \frac{\partial{\mathbf{u}}}{\partial t}&=&-\nabla \mathbf{u}\cdot \mathbf{u}-\nabla p+ \frac{1}{Re}\Delta \mathbf{u}+\mathbf{f}(t) \\ 
        \nabla \cdot \mathbf{u}&=&0.
    \end{eqnarray} 
    \end{subequations}
where $\mathbf{u}=(u,v)$ denotes the velocity field, $p$ is the pressure, and $Re$ is the Reynolds number based on the characteristic velocity $u_{\infty}$ and length scale $D$.
    
For later convenience, the governing equations \eqref{eq:NS} are written in the compact operator form 
\begin{equation}
\mathscr{E} \frac{\partial \mathbf{U}}{\partial t} = \mathscr{N}(\mathbf{U}, \mathbf{f}, Re),
\label{eq:NS_compact}
\end{equation}
where the state vector $\mathbf{U}$ is defined as
\[
\mathbf{U} =
\begin{pmatrix}
\mathbf{u} \\
p
\end{pmatrix},
\qquad
\mathscr{E} =
\begin{pmatrix}
\mathscr{I} & 0 \\
0 & 0
\end{pmatrix}.
\]
The nonlinear Navier--Stokes operator $\mathscr{N}$ is given by
\[
\mathscr{N}(\mathbf{U}, \mathbf{f}, Re) =
\begin{pmatrix}
- \nabla \mathbf{u}\cdot \mathbf{u} - \nabla p + \frac{1}{Re}\Delta \mathbf{u} + \mathbf{f}(t) \\
\nabla \cdot \mathbf{u}
\end{pmatrix}.
\]
Here, $\mathscr{E} = \mathscr{P}\mathscr{P}^{T}$, where $\mathscr{P}$ is the prolongation operator mapping the velocity field $\mathbf{u}$ to $(\mathbf{u},0)$, $\mathscr{P}^{T}$ is the corresponding restriction operator mapping $(\mathbf{u},p)$ to $\mathbf{u}$, and $\mathscr{I}$ denotes the identity operator. The steady (base) flow $\mathbf{U}_b$ is defined as an equilibrium solution of the unforced system, satisfying
\begin{equation}
\mathscr{N}(\mathbf{U}_b, \mathbf{0}, Re) = \mathbf{0}.
\label{eq:baseflow}
\end{equation}
In addition to the governing equations~\eqref{eq:NS_compact}, we introduce an observation model for partial measurements of the flow field. We assume that only limited flow information is available, as is typical in experiments, where measurements are obtained from a finite number of sensors, and denote these measurements by $\mathbf{M}$. These measurements are related to the full flow state through
\begin{equation}
\mathbf{M} = \mathscr{H}(\mathbf{U}),
\label{eq:measurement}
\end{equation}
where $\mathscr{H}$ denotes the observation operator.

\subsection{Hopf Bifurcation}

In the absence of forcing, i.e.\ for $\mathbf{f}(t)=\mathbf{0}$, and at sufficiently low Reynolds numbers, the flow past
two circular cylinders arranged in tandem is steady and symmetric, corresponding to the base flow $\mathbf{U}_b$ given by~\eqref{eq:baseflow}.
 As the Reynolds number increases beyond a
critical value $Re_c$, the steady base flow becomes globally unstable through a
Hopf bifurcation and self-sustained vortex shedding develops. The critical Reynolds
number $Re_c$ and the type of Hopf bifurcation (supercritical or subcritical)
depend on the spacing ratio $\gamma$ between the cylinders.

Mizushima and Suehiro~\cite{mizushima2005instability} reported a supercritical Hopf
bifurcation for small ($\gamma<3.4$) and large ($\gamma>5.5$) spacing ratios, and a
subcritical Hopf bifurcation for intermediate gaps ($3.4<\gamma<5.5$). In the
present work, we restrict our attention to the supercritical Hopf regime.

In the case of supercritical Hopf bifurcation, the steady base flow $\mathbf{U}_b$ loses stability when a pair of complex-conjugate eigenvalues $(\lambda,\lambda^{\star})$ of the generalized eigenvalue problem 
\begin{equation}
\mathscr{A}_b \hat{\mathbf{U}} = \lambda\, \mathscr{E}\hat{\mathbf{U}},
\label{eq:hopf_eigenmode}
\end{equation}
crosses the imaginary axis as the Reynolds number increases, while all remaining eigenvalues remain in the stable half-plane.
Here $\mathscr{A}_b$ denotes the Navier--Stokes operator linearized about $\mathbf{U}_b$ . At $Re=Re_c$, the leading eigenvalues satisfy
\begin{equation}
\lambda_0 = \imath \omega_0,
\label{eq:hopf_condition}
\end{equation}
    with $\omega_0>0$ denoting the bifurcation frequency and the corresponding eigenvector pair is referred to here as the \textit{critical global mode}. 
For $Re$ slightly above $Re_c$, the unforced dynamics lie on a two-dimensional invariant manifold which connects the unstable equilibrium (the base flow $\mathbf{U}_b$) and an attracting limit cycle~\cite{kuznetsov1998elements} associated
with periodic vortex shedding.

For large spacing ratios $\gamma>5.5$, this two-dimensional global instability is characterized by
vortex shedding both in the inter-cylinder gap and in the wake of the downstream
cylinder, corresponding to the co-shedding regime~\cite{carmo2010possible}. At higher
Reynolds numbers, typically $Re \gtrsim 180$, three-dimensional instabilities arise~\cite{carmo2010possible}, similarly to the single-cylinder case. 

\subsection{Problem Formulation}\label{sec:problemformulation}

\begin{figure*}
\centering
\includegraphics[width=1\textwidth]{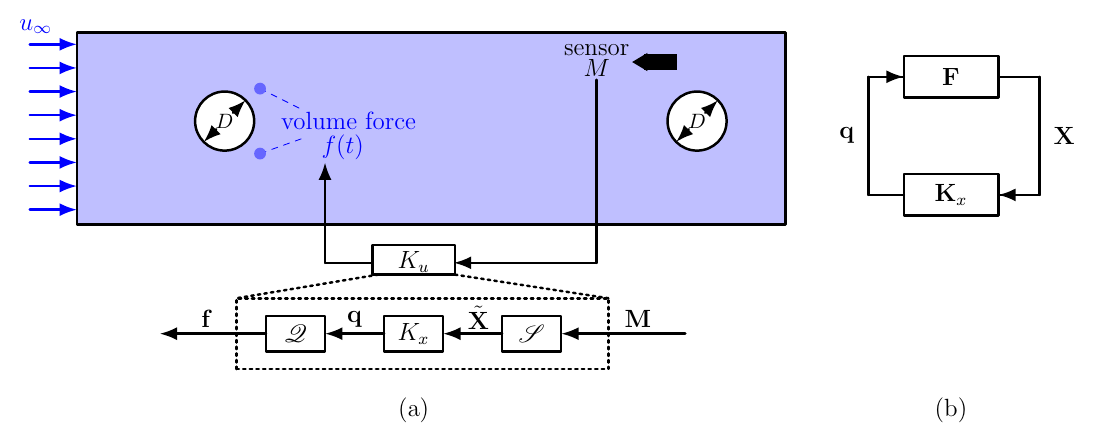}
\caption{Schematic of the closed-loop architecture implemented here to control the oscillating flow around tandem cylinders. (a) Output-feedback loop for the forced incompressible Navier-Stokes equations (full-order plant). (b) State-feedback loop for the reduced-order model.}
\label{fig:problemformulation}
\end{figure*}

The objective of this work is to suppress the two-dimensional global instability of the flow past tandem cylinders by designing a model-based closed-loop control law. 
We consider large spacing ratios for which the instability arises through a supercritical Hopf bifurcation, and Reynolds numbers sufficiently low for the flow to remain two-dimensional.

Control is applied through a time-varying volumetric forcing $\mathbf{f}(t)$ in the incompressible Navier--Stokes equations~\eqref{eq:NS_compact}. We assume that only partial measurements $\mathbf{M}$ of the flow state are available through the observation model~\eqref{eq:measurement}. The goal is therefore to construct an output-feedback law 
\begin{equation}\label{eq:feedback}
   \mathbf{f}\equiv \mathbf{K}_u(\mathbf{M})  
\end{equation}
that drives the system toward the steady base flow $\mathbf{U}_b$ from initial conditions spanning the relevant state space, from the vicinity of the equilibrium to the oscillating limit-cycle.

To build such a controller in a model-based manner, a fast evaluation of the system dynamics is required. However, the numerical approximation of~\eqref{eq:NS_compact} typically relies on a fine spatial discretization, resulting in a high-dimensional nonlinear model. Designing controllers directly for such systems is challenging, in particular for approaches based on online optimization. We therefore introduce a parametric reduced-order model
\begin{eqnarray}\label{eq:rom}
\frac{d\mathbf{X}}{dt}=\mathbf{F}(\mathbf{X},\mathbf{q},Re),
\end{eqnarray}
with a low-dimensional state $\mathbf{X}$ and input $\mathbf{q}$. The model is complemented by mappings $\mathscr{G}$ and $\mathscr{Q}$, which connect the reduced-order variables to the full-order state and forcing. In particular, $\mathbf{U}\approx \mathscr{G}(\mathbf{X})$ and $\mathbf{f}=\mathscr{Q}(\mathbf{q})$, where $\mathbf{U}$ denotes the solution of~\eqref{eq:NS_compact} under the forcing $\mathbf{f}$ and $\mathbf{X}$ is the corresponding solution of~\eqref{eq:rom}. The reduced-order model is sought to capture the relevant forced nonlinear dynamics across a range of parameter values.

Based on the reduced-order model~\eqref{eq:rom}, we design a state-feedback law
\begin{equation}\label{eq:statefeedback}
\mathbf{q}=\mathbf{K}_x(\mathbf{X},Re).
\end{equation}
To apply this controller to the full-order system using partial-state measurements, use an estimator $\mathscr{S}$ to reconstruct the reduced state
\begin{equation}\label{eq:state_estimation}
 \widetilde{\mathbf{X}}=\mathscr{S}(\mathbf{M}).   
\end{equation}
This way, we obtain the output-feedback law
\[
\mathbf{f}\equiv \mathbf{K}_u(\mathbf{M})
=\mathscr{Q}(\mathbf{K}_x(\widetilde{\mathbf{X}},Re)).
\]
for the full-order model. A schematic of the resulting closed-loop controller architecture is shown in Figure~\ref{fig:problemformulation}.

\section{Reduced-Order Model}\label{sec:WNA}
 
In the following, we outline the methodology for constructing a forced  reduced-order model~\eqref{eq:rom} of the incompressible Navier--Stokes equations~\eqref{eq:NS_compact}. We build on the weakly nonlinear analysis of \citet{sipp2007global,sipp2012open} to obtain a two-dimensional forced Stuart--Landau model. In contrast to the formulation in \citet{sipp2012open}, where the reduced-order input is assumed to be constant, we allow a time-dependent input $\mathbf{q}(t)$ in~\eqref{eq:rom}. This enables the design of closed-loop feedback controllers at the slow time scale. Although we apply the methodology here to the flow past tandem cylinders, it is applicable to any flow governed by~\eqref{eq:NS_compact} that loses stability through a supercritical Hopf bifurcation.

\subsection{Weakly Nonlinear Analysis}

Following the classical weakly nonlinear analysis for flows governed by the incompressible Navier--Stokes equations near the onset of a supercritical Hopf bifurcation from~\citet{sipp2007global}, we consider Reynolds numbers slightly above the critical value $Re_c$ and introduce the small bifurcation parameter
\begin{equation}\label{eq:reynolds}
\epsilon := Re_c^{-1}-Re^{-1}, \qquad 0<\epsilon\ll 1.
\end{equation}
The analysis is based on a separation of time scales, with a fast time $t$ associated with the oscillatory dynamics of the critical global mode and a slow time $\tau=\epsilon t$ governing the modulation of the oscillation amplitude.
The flow state $\mathbf{U}=(\mathbf{u},p)^T$ is expanded asymptotically as
\begin{equation}\label{eq:expansion}
\mathbf{U}(t) \approx \mathbf{U}_0
+ \sqrt{\epsilon}\,\mathbf{U}_1(t,\tau)
+ \epsilon\,\mathbf{U}_2(t,\tau)
+ \epsilon\sqrt{\epsilon}\,\mathbf{U}_3(t,\tau)
+ \cdots,
\end{equation}
about the steady unforced base flow $\mathbf{U}_0=(\mathbf{u}_0,p_0)^T$ at $Re_c$.
Here, the approximate equality signifies the fact that asymptotic expansions are not necessarily convergent over the whole domain \cite{vandyke75perturbation}. 

We next introduce a volumetric forcing  of the form
\begin{equation}\label{eq:forcing}
\mathbf{f}(t,\tau) = E'(\tau)\,e^{\mathrm{i}\omega_f t}\,\mathbf{f}_E + \mathrm{c.c.},
\end{equation}
where $\mathbf{f}_E$ is the complex spatial structure of the forcing, and $\omega_f$ is the forcing frequency on the fast time scale. We allow the \emph{complex} forcing amplitude $E'(\tau)$ to vary on the slow time scale, generalizing the approach form~\cite{sipp2012open} which assumes constant $E'$. Writing $E'(\tau)=|E'(\tau)|e^{\mathrm{i}\phi_E(\tau)}$
shows that both the magnitude of the forcing amplitude $|E'(\tau)|$ and the phase $\phi_E(\tau)$ evolve
on the slow time scale $\tau$. Consequently, the 
instantaneous frequency of the forcing also varies slowly. 
Substituting the expansion \eqref{eq:expansion} and forcing \eqref{eq:forcing} into the governing equations \eqref{eq:NS_compact} and collecting terms at successive orders in $\sqrt{\epsilon}$ yields a hierarchy of linear inhomogeneous problems for $\mathbf{U}_1,\mathbf{U}_2,\mathbf{U}_3,\ldots$, which are solved sequentially.

The forcing amplitude $E'(\tau)$ is assumed to scale with $\epsilon$ as
\begin{equation}\label{eq:forcing_scaling}
E'(\tau) =\sqrt{\epsilon}^{\,i}\,E(\tau),
\end{equation}
where the exponent $i$ determines the order at which the forcing enters this hierarchy. Following \citet{sipp2012open}, different values of $i$ are chosen for different forcing-frequency classes so that the forcing does not lead to degenerate operator equations at $\mathcal{O}(\sqrt{\epsilon})$ and $\mathcal{O}(\epsilon)$.

Here we proceed with the case of resonant forcing frequencies near $\omega_0$, since it allows full stabilization of the flow, as explained in Appendix~\ref{sec:stabilization}.
Other frequency cases are discussed in Appendix~\ref{sec:otherappendix} and~\ref{sec:nonresonant}.

\subsection{Resonant Forcing}
Following the scaling suggested by~\citet{fauve2009} and \citet{sipp2012open}, we consider the forcing amplitude 
\begin{equation}\label{eq:scaling_resonant}
E'(\tau):=\sqrt{\epsilon}^{\,3}E(\tau)
\end{equation}
when $\omega_f=\omega_0+\epsilon \Omega$ for some frequency $\Omega$.

Solving the corresponding equations at orders $\sqrt{\epsilon}^{\,0}$, $\sqrt{\epsilon}^{\,1}$, $\sqrt{\epsilon}^{\,2}$, and $\sqrt{\epsilon}^{\,3}$, $\mathbf{U}$ is obtained in the form
\begin{equation}\label{eq:solution_wna1}
\begin{split}
\mathbf{U}\approx{}& \mathbf{U}_0
+ \sqrt{\epsilon}\big(Ae^{\imath \omega_0 t}\mathbf{U}_1^A + c.c.\big) \\
&+ \epsilon\big(\mathbf{U}_2^1 + |A|^2 \mathbf{U}_2^{|A|^2}
+ \big(A^2 e^{\imath 2\omega_0 t}\mathbf{U}_2^{A^2} + c.c.\big)\big) \\
&+ \sqrt{\epsilon}^{\,3}\big(
A e^{\imath \omega_0 t}\mathbf{U}_3^A
+ A|A|^2 e^{\imath \omega_0 t}\mathbf{U}_3^{A|A|^2} \\
&\qquad\qquad
+ E e^{\imath (\omega_0+\epsilon\Omega)t}\mathbf{U}_3^E
+ c.c.\big)
+ \ldots
\end{split}
\end{equation}
 To ensure that the solution~\eqref{eq:solution_wna1} remains bounded in time, and hence, retain consistency of the asymptotic expansion (see \ref{sec:resonantappendix}), the evolution of the complex amplitude $A$, also referred to as the global mode amplitude, needs to satisfy the Stuart-Landau equation
\begin{equation}\label{eq:SL_resonance}
    \phantom{.}\frac{d A}{d \tau}=a_0 A - a_1 A \lvert A \rvert^2 +a_2 e^{\imath \epsilon \Omega t}E. 
\end{equation}
This can be rewritten with respect to the fast timescale $t$ as 
\begin{equation}\label{eq:SL_resonance_t}
   \frac{d A}{d t}=\epsilon a_0 A - \epsilon a_1 A \lvert A \rvert^2 + \epsilon a_2e^{\imath \epsilon \Omega t} E. 
   \end{equation}
The coefficients $a_0$, $a_1$, and $a_2$ are determined by orthogonality conditions that are given in Appendix~\ref{sec:resonantappendix}, where detailed derivation of the Stuart--Landau equation is also given.

We briefly comment on the physical interpretation of the weakly nonlinear expansion~\eqref{eq:solution_wna1} and the Stuart-Landau model~\eqref{eq:SL_resonance_t}. 

In the unforced case ($E=0$) and for $Re>Re_c$, the Stuart--Landau equation admits two lower-dimensional invariant sets: the unstable equilibrium $A=0$, and a stable limit cycle with constant magnitude
\begin{equation}\label{eq:LC_magnitude}
 |A|_{LC}=\sqrt{\frac{\Re(a_0)}{\Re(a_1)}}   
\end{equation}
and constant phase velocity. This limit-cycle oscillation of the full flow field is captured by the expansion~\eqref{eq:solution_wna1}. 
As the slowly varying parameters in this expansion evolve according to the Stuart-Landau dynamics, the corresponding flow field evolves on a two-dimensional manifold connecting the aforementioned invariant sets and governing the long-time behavior of the flow.

The steady base flow is approximated from~\eqref{eq:solution_wna1} as
$\mathbf{U}_b \approx \mathbf{U}_0+\epsilon\,\mathbf{U}_2^{1}$, where $\mathbf{U}_2^{1}$ is the base flow correction for $Re>Re_c$. 
As the flow evolves toward the limit cycle, the mean (time-averaged) flow departs slowly from the base flow through the $\mathcal{O}(\epsilon|A|^2)$ contribution
$\epsilon |A|^2\,\mathbf{U}_2^{|A|^2}$.
In analogy with the terminology used in~\citet{noack2003hierarchy}, $\mathbf{U}_2^{|A|^2}$ is referred to as the \emph{shift mode}.

The unsteady component oscillating at the fundamental frequency is given by the terms in~\eqref{eq:solution_wna1} proportional to $e^{\mathrm{i}\omega_0 t}$.
The linear contribution,
$\sqrt{\epsilon}\,A e^{\mathrm{i}\omega_0 t}\mathbf{U}_1^A
+\sqrt{\epsilon}^{\,3}\,A e^{\mathrm{i}\omega_0 t}\mathbf{U}_3^A
+\mathrm{c.c.}$,
represents the approximation of the unstable eigenmode of the linearized system about $\mathbf{U}_b$ at the considered Reynolds number: $\mathbf{U}_1^A$ corresponds to the critical global mode, while $\mathbf{U}_3^A$ accounts for its correction for $Re>Re_c$.
The nonlinear correction 
$\sqrt{\epsilon}^{\,3}\,A|A|^2 e^{\mathrm{i}\omega_0 t}\mathbf{U}_3^{A|A|^2}
+\mathrm{c.c.}$ 
describes the slow, amplitude-dependent deformation of the fundamental mode as the trajectory evolves along the attracting manifold toward the limit cycle.
The fundamental frequency is likewise detuned for $Re>Re_c$: it differs from $\omega_0$ by an $\mathcal{O}(\epsilon)$ correction near the equilibrium and varies slowly with the amplitude until it approaches a constant value on the limit cycle predicted by the Stuart--Landau phase dynamics~\cite{sipp2007global}.
The \emph{second-harmonic} contribution
$\epsilon A^2 e^{\mathrm{i}2\omega_0 t}\mathbf{U}_2^{A^2}+\mathrm{c.c.}$
oscillates at twice the fundamental frequency.

The leading forcing-induced contribution enters~\eqref{eq:solution_wna1} at $\mathcal{O}(\sqrt{\epsilon}^{\,3})$ through the term proportional to $E$. This term perturbs the flow away from the invariant manifold of the unforced dynamics. Still, the forced flow remains close to this manifold due to its string attractivity.

\subsection{Forced Stuart-Landau Model} \label{sec:forcedSL}

In real coordinates $\mathbf{X}=[\Re(A), \Im(A)]^T$ with input $\mathbf{q}=[\Re(E), \Im(E)]^T$, the Stuart-Landau equation~\eqref{eq:SL_resonance_t}
takes the form~\eqref{eq:rom} of a parametric reduced-order model of the incompressible Navier-Stokes equations~\eqref{eq:NS_compact}. The model captures the nonlinear forced dynamics, incorporates time-dependent forcing, and depends explicitly on the Reynolds number. 
Moreover, the weakly nonlinear expansion~\eqref{eq:solution_wna1}, together with the forcing definitions~\eqref{eq:forcing} and~\eqref{eq:scaling_resonant}, provides explicit mappings from the reduced variables to the full-order quantities. In particular, we obtain $
\mathscr{G}:\mathbf{X}\mapsto \mathbf{U}\approx \mathscr{G}(\mathbf{X})$ and $\mathscr{Q}:\mathbf{q}\mapsto \mathbf{f}= \mathscr{Q}(\mathbf{q})$,
which map the reduced-model state and input spaces to the corresponding spaces of the full-order system~\eqref{eq:NS_compact}. 
The difference between the full state and its reconstruction from the reduced model is captured by the approximation error
\begin{equation}\label{eq:approximationerror}
    \mathbf e\equiv(\mathbf{e}_{\mathbf{u}}, e_p):=\mathbf U-\mathscr{G}(\mathbf{X}).
\end{equation}
These mappings enable the construction of the estimator~\eqref{eq:state_estimation} and the deployment of the reduced-order-model controller on the full-order system, as described next.

\section{Closed-Loop Control}\label{sec:control}
In the following, we design an output-feedback law~\eqref{eq:feedback} for the Navier--Stokes equations~\eqref{eq:NS_compact}, assuming access to the partial-state measurements~\eqref{eq:measurement} and using the reduced-order model derived in Sec.~\ref{sec:WNA}. 
This is implemented in three successive steps, consistent with the architecture summarized in Sec.~\ref{sec:problemformulation}. 
First, the reduced state is reconstructed from the available velocity measurements using the estimator~\eqref{eq:state_estimation}. 
Second, the resulting state estimate is supplied to the reduced-model controller~\eqref{eq:statefeedback}, which computes the control input  $\mathbf{q}=[\Re(E), \Im(E)]^T$ in the reduced-order dynamics~\eqref{eq:SL_resonance_t}. These two steps are described in 
Sections~\ref{sec:estimation}  and \ref{sec:mpc}.
Third, this control input is translated into the full-order forcing $\mathbf{f}(t,\tau)$ via mapping $\mathscr{Q}$. The combination of these three steps yields the output-feedback forcing $\mathbf{f}\equiv \mathbf{K}_u(\mathbf{M})$ applied to the Navier--Stokes equations.

We consider the forcing~\eqref{eq:forcing} with frequency $\omega_f=\omega_0$ for two reasons. First, for $\omega_f=\omega_0$ the control amplitude $E$ enters the Stuart--Landau equation~\eqref{eq:SL_resonance_t} as an additive input, which is not the case for other frequency classes, except $\omega_f=\omega_0/2$ (see Appendix~\ref{sec:stabilization}). This allows the controller to stabilize the model while driving $E$ to zero. When both $A$ and $E$ converge to zero, the flow approaches the equilibrium base state $\mathbf{U}_b\approx \mathbf{U}_0+\epsilon\mathbf{U}_2^1$ according to~\eqref{eq:solution_wna1}, and the unsteadiness is suppressed. Second, this forcing case facilitates the characterization of the spatial forcing structure $\mathbf{f}_E$ that maximizes control efficiency, as shown in Section~\ref{sec:optimalforcing}.

\subsection{Reduced state estimation from measurements} \label{sec:estimation}

We reconstruct the complex amplitude $A$ from the available flow measurements $\mathbf{M}$. Since the reduced-order dynamics are two-dimensional and we assume noise-free measurements, we employ an \emph{instantaneous} (static) reconstruction of the reduced state from  measurements. In particular, provided that the measurements contain at least two independent scalar quantities (so that the real and imaginary parts of $A$ can be uniquely determined), a dynamic estimator such as an observer or a Kalman filter is typically not required.
\color{black}
In practice, the state reconstructed from measurements, denoted by $\widetilde{A}$, generally differs from the reduced-order state $A$ predicted by the surrogate dynamics due to the approximation errors introduced by the model. 

We consider two measurement settings: (i) full-field velocity measurements and (ii) pointwise velocity  measurements at a finite number of sensor locations.

If the whole velocity field is measured, i.e. $\mathbf{M}=(\mathbf{u},0)^T$, we use the reduced-to-full state mapping
\begin{align}\label{eq:Gmap}
\mathscr G(\mathbf X):=\mathbf{U}_0
+&\sqrt{\epsilon}\big(Ae^{\imath \omega_0t}\mathbf{U}_1^A+c.c.\big) \nonumber \\
+&\epsilon(\mathbf{U}_2^1+ \lvert A \rvert^2 \mathbf{U}_2^{\lvert A \rvert^2} + \big(A^2 e^{\imath 2\omega_0t}\mathbf{U}_2^{A^2}+ c.c.\big)\big)
\end{align}
which retains the terms up to second order from~\eqref{eq:solution_wna1}.
The corresponding dual mapping $\mathscr{S}$ from the full to the reduced state is obtained by projecting~\eqref{eq:Gmap} onto the velocity component of the adjoint global mode $\mathbf{u}_1^{A \star}$ at $Re_c$, i.e., by
\begin{equation}\label{eq:Atilde}
    \mathscr{S}(\mathbf{M}) \equiv  \tilde{A}:= \frac{1}{\sqrt{\epsilon}}\langle \mathbf{u}_1^{A \star}, \mathbf{u}-\mathbf{u}_0\rangle e^{-\imath \omega_0 t},
\end{equation}
where $\mathbf{u}_1^{A \star}$ is normalized such that, $\langle \mathbf{u}_1^{A \star}, \mathbf{u}_1^A  \rangle=1$. Owing to the specific symmetry properties in the present test case, all contributions at $\mathcal{O}(\epsilon)$ vanish under this projection. Hence, we obtain an explicit expression for $\tilde{A}$. A detailed discussion of the symmetry arguments leading to this cancellation is provided in Section~\ref{sec:results}.

We next consider measurements consisting of $x$- and/or $y$-velocity components at sensor locations $\mathbf{x}_i$, which we write as
$\mathbf{M}=\{(u(\mathbf{x}_i), v(\mathbf{x}_i))^{T}\}_{i=1}^{n}$.
In this case, reconstructing $A$ requires at least two independent scalar measurements. We obtain $\widetilde{A}$ by using a linearized measurement model that retains only the terms that are linear in $A$. This approximation is justified when the contributions of $\mathbf{U}_2^{|A|^2}$ and $\mathbf{U}_2^{A^2}$ are small at the sensor locations. In Section~\ref{sec:results} we show that this is the case for sensors placed close to the upstream cylinder. The resulting reconstruction is written in terms of the real and imaginary parts as

\begin{equation}\label{eq:measurements}
\begin{aligned}
\mathscr{S}(\mathbf{M}) & \equiv \begin{bmatrix}
 \Re (\tilde{{A}}_p) \\
    \Im (\tilde{{A}}_p)
\end{bmatrix}:=
\frac{1}{2 \sqrt{\epsilon}} e^{-\imath \omega_0t} {\underbrace{\begin{bmatrix}
    \Re (u_1^A(\mathbf{x}_1)) & -\Im (u_1^A(\mathbf{x}_1)) \\
     \Re (v_1^A(\mathbf{x}_1)) & -\Im (v_1^A(\mathbf{x}_1)) \\
    \vdots & \vdots \\
        \Re (u_1^A(\mathbf{x}_n)) & -\Im (u_1^A(\mathbf{x}_n)) \\
        \Re (v_1^A(\mathbf{x}_n)) & -\Im (v_1^A(\mathbf{x}_n))
\end{bmatrix}}_{\mathbf{G}}}^{\dagger}\begin{bmatrix}
u(\mathbf{x}_1)- u_0(\mathbf{x}_1)-\epsilon u_2^1(\mathbf{x}_1) \\
v(\mathbf{x}_1)- v_0(\mathbf{x}_1)-\epsilon v_2^1(\mathbf{x}_1) \\
 \vdots \\
  u(\mathbf{x}_n)-u_0(\mathbf{x}_n)-\epsilon u_2^1(\mathbf{x}_n) \\
    v(\mathbf{x}_n)-v_0(\mathbf{x}_n)-\epsilon v_2^1(\mathbf{x}_n)
\end{bmatrix}. 
\end{aligned}
\end{equation}

\subsection{Model Predictive Control}\label{sec:mpc}

We design a model predictive controller for the reduced-order model~\eqref{eq:SL_resonance_t}, which seeks to optimally steer the amplitude of the global mode $A$ to zero while incorporating suitable constraints for the forcing amplitude $E$. The latter include soft constraints on the magnitude of the forcing and its rate of change, which are introduced as penalty terms in the objective function of the optimization problem. This enables us to to bring the control input $E$ towards zero while naturally complying with our modeling requirement of small temporal gradients of $E$, to be consistent with the separation of timescales assumed in the derivation of the reduced-order model. 

To determine the inputs of the model predictive controller, we consider a discretized version of the forced Stuart-Landau model~\eqref{eq:SL_resonance_t} with sampling period $\Delta t_{MPC}$ and zero-order hold of the complex input amplitude $E$. We assume that $\Delta t_{MPC}$ is sufficiently large compared to the fast timescale, so that potential jumps of $E$ across the sampling times have no significant impact on the validity of the weakly nonlinear analysis. 

MPC relies on recursively solving a finite-horizon optimal control problem at each time instant and keeping only the first input of the resulting optimization problem. Considering a horizon of $m$ time steps and using the notation $\mathbf{X}_j$ and $\mathbf{q}_j$ for the state and input of the time-discretized Stuart-Landau model at time $t_j=j\Delta t_{MPC}$, we seek to minimize the quadratic cost function

\begin{equation}\label{eq:costfunction}
    J(\mathbf{X}_{[j:j+m]},\mathbf{q}_{[j:j+m-1]}):= \sum_{k=1}^m\| \mathbf{X}_{j+k}-\mathbf{X}^{\ast} \|^2_{\mathbf{Q}}+ \sum_{k=0}^{m-1}\big( \| \mathbf{q}_{j+k}\|^2_{\mathbf{R_u}}+ \| \Delta \mathbf{q}_{j+k}\|^2_{\mathbf{R}_{\Delta \mathbf{u}}}\big)
\end{equation}

subject to the dynamics of the system. Here $\mathbf{q}_{[j:j+m-1]}:=(\mathbf{q}_j,..., \mathbf{q}_{j+m-1})$ denotes the sequence of control actions applied over the horizon and $\mathbf{X}_{[j:j+m]}:=(\mathbf{X}_j,...,\mathbf{X}_{j+m})$ denotes the corresponding state sequence of the system. 

Since the model predictive controller is used for the output feedback of the full-order plant, its initial state satisfies the constraint $\mathbf{X}_j=\widetilde{\mathbf{X}}_j$, where the reduced state $\widetilde{\mathbf{X}}_j$ 
is estimated from the measurement 
$\mathbf M(t_j)$

of the flow field at time $t_j$ either  via or \eqref{eq:Atilde} or \eqref{eq:measurements}.

\begin{figure}
\centering
\includegraphics[width=0.8\textwidth]{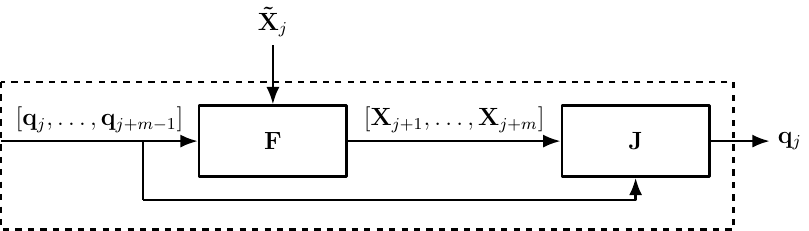}
\caption{Schematic of the model-predictive controller. At each time step $t_j$, the future states $\mathbf{X}_{j+1},...,\mathbf{X}_{j+m}$ are predicted by the surrogate model $\mathbf{F}$ 
with the initial state $\mathbf{X}_j$ set equal to its estimate $\widetilde{\mathbf{X}}_j$ from the flow measurements. The inputs are determined by minimizing the cost function $J$. The first input $\mathbf{q}_j$ is applied to the
system and the same process is repeated at the next time step.} \label{fig:MPC}
\end{figure}

The future states $\mathbf{X}_{j+1},...,\mathbf{X}_{j+m}$ are predicted by the time-discretized version of the Stuart-Landau model~\eqref{eq:SL_resonance_t} and the inputs are selected in such a way that they minimize the corresponding cost \eqref{eq:costfunction}. Then the first input $\mathbf q_j$ is applied to the system and the same process is repeated recursively. The overall procedure is illustrated schematically in Figure~\ref{fig:MPC}.   Each term in~\eqref{eq:costfunction} 
is a quadratic semi-norm of a vector $\mathbf{x}$, denoted by $\| \mathbf{x} \|^2_{\mathbf{S}} :=\mathbf{x}^T\mathbf{S}\mathbf{x}$ for some positive semi-definite matrix $\mathbf{S}$.
The cost function $J$ 
penalizes the deviations of the
predicted state $\mathbf{X}_{j+k}$ from the reference state $\mathbf{X}^{\ast}= [0,0]$ as well as the magnitude 
of the input $\mathbf{q}_{j+k}$ and its increments $\Delta \mathbf{q}_{j+k}=\mathbf{q}_{j+k}-\mathbf{q}_{j+k-1}$, with the first computed using the previously selected input $\mathbf{q}_{j-1}$ of the MPC recursion. 
The corresponding weight matrix $\mathbf{Q}$ is positive definite, while $\mathbf{R_{u}}$ and 
$\mathbf{R_{\mathnormal{\Delta} u}}$ are positive semi-definite.

\subsection{Optimal Forcing Structure}\label{sec:optimalforcing}
The choice of the forcing structure $\mathbf{f}_E$ directly affects control efficiency through the coefficient $a_2$ in~\eqref{eq:SL_resonance_t}. In particular, due to the structure of the forced Stuart--Landau dynamics, which are fully actuated with respect to $E$, the higher the magnitude $|a_2|$, the lower the magnitude of the forcing amplitude $|E|$ required to bring $|A|$ to zero. To make this argument precise, let $t\mapsto E_\alpha(t)$ be a (potentially optimal) control trajectory that stabilizes the system, which is subject to the forcing structure $\mathbf f_{E,\alpha}$, and hence, has a forcing coefficient  $a_{2,\alpha}$. When the system is subject to another forcing structure $\mathbf f_{E,\beta}$ with a larger coefficient $a_{2,\beta}$, the input $t\mapsto  E_\beta(t)= a_{2,\alpha}/a_{2,\beta}E_\alpha(t)$ will have a smaller magnitude and yield the exact same state trajectory. This means that we can always achieve the same control objective as in the first case with a smaller control effort. This is also consistent with the MPC cost in \eqref{eq:costfunction}, which will be respectively lower when the weight matrices $\mathbf{R_{u}}$ and $\mathbf{R_{\mathnormal{\Delta} u}}$ are multiples of the identity matrix.

We thus seek a forcing structure $\mathbf{f}_E$ of unit energy $\langle \mathbf{f}_E, \mathbf{f}_E \rangle=1$ that gives the highest magnitude of $|a_2|$.
For the resonant case where $\omega_f \approx \omega_0$, $|a_2|$ is maximized when the forcing is aligned with the velocity component of the adjoint global mode $\mathbf{u}_1^{A \star}$ at $Re_c$. This follows directly from the definition of the coefficient $a_2$
 (cf.~\eqref{eq:a2} in Appendix~\ref{sec:WNAappendix}) and is consistent with the findings of \citet{sipp2012open}. 
 
 Thus, we obtain an optimal forcing structure $\mathbf{f}_E$ by normalizing $\mathbf{u}_1^{A \star}$, i.e., by selecting
\begin{equation}\label{eq:optimal_forcing}
 \mathbf{f}_E:=\frac{c\mathbf{u}_1^{A \star}} {|c|\sqrt{\langle \mathbf{u}_1^{A \star}, \mathbf{u}_1^{A \star} \rangle}}.  
\end{equation}
where $c$ is a complex constant.

Similar observations have been made using resolvent analysis for the flow around a single cylinder, where the optimal forcing on the linearized Navier--Stokes equations is obtained from the singular value decomposition (SVD) of the resolvent operator. It has been shown in~\citet{symon2018non} and ~\citet{jin2021energy} that the forcing producing the maximum energy gain occurs at the resonant frequency and has the same spatial structure as the adjoint mode. These results have been leveraged for the control of the single-cylinder wake in~\citet{jin2022resolvent} and~\citet{ma2024suppression}.
\color{black}

Applying a widely spatially distributed forcing is unrealistic in practice. We therefore also consider a more physically realizable forcing structure concentrated on a small domain $\Omega_s$. Again, the structure $\mathbf{f}_E$ which maximizes $|a_2|$ is determined by the restriction of $\mathbf{u}_1^{A \star}$ on $\Omega_s$ 
For a sufficiently small domain, we can assume that $\mathbf{u}_1^{A \star}(\mathbf{x})$ is approximately constant on $\Omega_s$. 
Thus, we may choose any point $\mathbf{x}_i\in\Omega_{s}$ and consider the forcing structure 
\begin{equation}\label{eq:localforcing}
  \mathbf{f}_E:=\frac{1}{\sqrt{{\rm vol}(\Omega_{s})}}\frac{\mathbf{u}_1^{A \star}(\mathbf{x}_i)}{||\mathbf{u}_1^{A \star}(\mathbf{x}_i)||}\chi_{\Omega_{s}}.  
\end{equation}
Here $\chi_{\Omega_{s}}$ denotes the indicator function of a set $\Omega_{s}\subset\mathbb R^2$, which takes the value $1$ on $\Omega_{s}$ and $0$ outside. 
It follows that $|a_2|=||\mathbf{u}_1^{A \star}(\mathbf{x}_i)|| \sqrt{\rm vol(\Omega_s)}$ for the uniform, normalized $\mathbf{f}_E$~\eqref{eq:localforcing}, where $ ||\mathbf{u}_1^{A \star}(\mathbf{x}_i)||=\sqrt{|u_1^{A \star}(\mathbf{x}_i)|^2+|v_1^{A \star}(\mathbf{x}_i)|^2}$ is the norm of $\mathbf{u}_1^{A \star}$ evaluated at the spatial location $\mathbf{x}_i$. This indicates that to maximize $|a_2|$, the volume force should be applied at the location $\mathbf{x}_i$ where the norm of the adjoint mode is the highest.

\section{Numerical Setup and Results}\label{sec:results}

In this section, we exploit our approach to numerically obtain a parametric reduced-order model and design a closed-loop controller for the flow around two cylinders in tandem configuration. We choose a spacing ratio of $\gamma=8$. This is large enough for the flow to exhibit supercritical Hopf bifurcation~\cite{mizushima2005instability}, and thus comply with the assumptions of our methodology. On the other hand, the gap is small enough for the wake from the upstream cylinder to interact strongly with the downstream cylinder, inducing high unsteady loads on the downstream cylinder~\cite{liu2024primary}, and making it a relevant study case for load alleviation control purposes. 
To showcase the parametric adaptability of our reduced model and control design, we consider multiple values of $Re$, specifically $Re=50,60,70$ and $80$.  

\subsection{Numerical Setup}

Here we present the computational mesh and numerical setup used for the weakly nonlinear analysis and for the direct numerical simulation (DNS) of the incompressible Navier--Stokes equations~\eqref{eq:NS} for the tandem-cylinder configuration. The DNS provides the time-resolved flow field, which serves as the ``ground truth'' used to validate the reduced-order model and to extract the measurement signals $\mathbf{M}$ employed in the feedback loop, either as full-field velocity data or as pointwise velocity samples.

\subsubsection{Computational Domain and Mesh}\label{sec:computationaldomain}

\begin{figure}
\centering
\includegraphics[width=0.8\textwidth, trim = 0pt 10pt 0pt 0pt, clip=true]{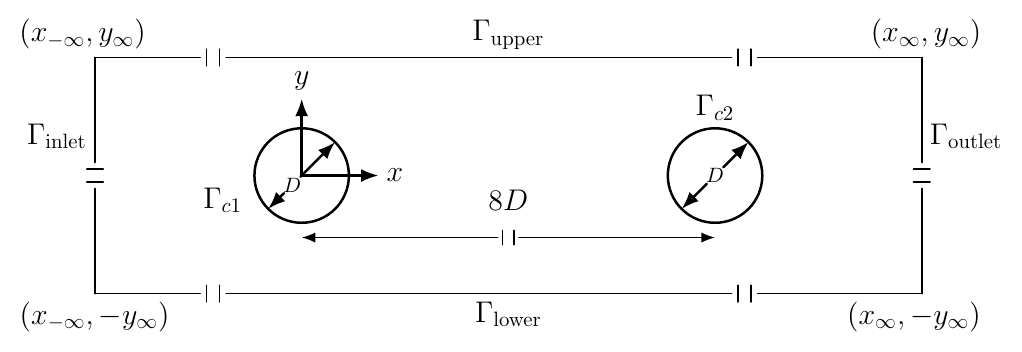}
\caption{Schematic defining the computational domain. 
The streamwise and transverse coordinates 
$x_{-\infty}$/$x_{\infty}$ and $-y_{\infty}$/$y_{\infty}$, determine respectively the location of the inlet/outlet and lower/upper boundaries. The primary flow direction is from left to right.
}\label{fig:mesh}
\end{figure}

The computational domain is shown in Figure~\ref{fig:mesh}, where the origin of the Cartesian coordinate system $\mathbf{x}=(x,y)$ is placed at the center of the upstream cylinder. Both cylinders are of diameter $D=1$ and the spacing ratio is $\gamma=8$. The domain extends to $x_{-\infty}=-60$ in the upstream direction, $x_{+\infty}=200$ in the downstream direction, and from $y_{-\infty}=-30$ to $y_{+\infty}=30$ in the transverse direction. 
We apply Dirichlet boundary conditions $\mathbf{u}=(1,0)$ on the inlet $\Gamma_{\rm inlet}$, no-slip boundary conditions $\mathbf{u}=(0,0)$ on the cylinder boundaries $\Gamma_{\rm c1}$ and $\Gamma_{\rm c2}$, the standard freestream boundary condition $(p-Re^{-1}\partial{u}/\partial{x}=0, \partial{v}/\partial{x}=0)$ at the outlet $\Gamma_{\rm outlet}$, and the symmetric boundary conditions $(\partial{u}/\partial{y}=0, v=0)$ on the upper and lower boundary, $\Gamma_{\rm upper}$ and $\Gamma_{\rm lower}$.

The incompressible Navier-Stokes equations are spatially discretized via the Finite Element Method (FEM) using the FreeFEM++ software~\cite{MR3043640}. We choose Taylor–Hood elements $P2$ for velocity and $P1$ for the pressure. 

Different meshes were used to check the convergence of the parameters of the weakly nonlinear analysis in Appendix~\ref{Meshconvergence}. In the rest of this paper, we use the mesh denoted with $M2$ to present our results.

The mesh is unstructured and the gridpoint distribution is determined by the automated mesh adaptation method AdaptMesh in FreeFEM++, which relies on the Delaunay-Voronoi algorithm. The mesh resolution is adapted to the base flow and the structure of the direct and adjoint global modes. The mesh has a total of $N_t=359028$ triangles and $N_{\rm dof}=1619133$ degrees of freedom.

\subsubsection{Numerical implementation of the weakly nonlinear analysis}
\label{ImplementationWNA}

To obtain the Stuart--Landau model~\eqref{eq:SL_resonance_t}, we solve the weakly nonlinear expansion up to order $\epsilon$, and determine the model coefficients at order $\epsilon^{3/2}$ from the corresponding compatibility conditions, following the procedure outlined in Appendix~\ref{sec:resonantappendix}.

Solving the equations at the first two orders of the expansion requires the base flow $\mathbf{U}_0$ at $Re_c$, together with the corresponding marginally stable eigenvalue $\lambda_0=\sigma_0 \pm \imath \omega_0$ (with $\sigma_0 \approx 0$) and eigenvector $\mathbf{U}_1^A$. The nonlinear steady Navier--Stokes equations~\eqref{eq:base} are solved using a Newton iteration, while the eigenvalue problem~\eqref{eq:eigenvalue} is solved using the shift-and-invert Arnoldi method as implemented in ARPACK~\cite{lehoucq1998arpack}.

Since $Re_c$ is not known a priori, this procedure is performed iteratively. Starting from an initial guess for $Re_c$, we compute the base flow and solve the corresponding eigenvalue problem. The process is repeated until $\sigma_0$ is sufficiently small.

To invert the $N_{\rm dof}\times N_{\rm dof}$ matrices of the discretized operator equations of the weakly nonlinear analysis, we use the UMFPACK library, which relies on a sparse direct LU solver~\cite{davis2004algorithm}.

\subsubsection{Direct Numerical Simulation}\label{sec:DNS}
 
The direct numerical simulation of the unsteady incompressible Navier-Stokes equations~\eqref{eq:NS_compact} marches the velocity-pressure field $\mathbf{U}$ in time. We consider the perturbation form  
\begin{equation}\label{eq:NS_perturbation}
   \mathscr{E} \frac{\partial{\mathbf{U}'}}{\partial t}=
   \mathscr{A} \mathbf{U}' -\mathcal{P}\big(
 \nabla \mathbf{u}'\cdot \mathbf{u}'-\mathbf{f}  \big)
\end{equation}
of the Navier-Stokes equations, where $\mathbf{U}'=(\mathbf{u}', p')=\mathbf{U}-\mathbf{U}_b$ contains the velocity and pressure components of the perturbation around the steady solution $\mathbf{U}_b$ (base flow) at some $Re$. First, we calculate the base flow $\mathbf{U}_b$ using the iterative Newton method as described in Section~\ref{ImplementationWNA}, and then we march $\mathbf{U}'$ in time.

For the time discretization of~\eqref{eq:NS_perturbation}, we use a second-order semi-implicit backward-finite-difference scheme, which gives at time $t_{i+1}=(i+1) \Delta t_{DNS}$ the field 
\begin{equation}\label{eq:NS_perturbation_td}
\begin{aligned}
 \mathbf{U}'_{i+1}=&\Big(\frac{3}{2 \Delta t_{DNS}} \mathscr{E}-\mathscr{A}\Big)^{-1} \Big(\frac{2}{\Delta t_{DNS}} \mathscr{E}\mathbf{U}'_{i}-\frac{1}{2\Delta t_{DNS}} \mathscr{E}\mathbf{U}'_{i-1}-
 \\
 &\big.\mathcal{P}\big( 2\nabla \mathbf{u}'_{i}\cdot \mathbf{u}'_{i}-\nabla \mathbf{u}'_{i-1}\cdot \mathbf{u}'_{i-1}-2\mathbf{f}_{i}+\mathbf{f}_{i-1}
   \big)\Big).
\end{aligned}
\end{equation}
We use the spatial discretization and finite elements from Section~\ref{sec:computationaldomain}, and use different time steps $\Delta t_{DNS}$ depending on the Reynolds number. For the unforced DNS at $Re\leq 70$, $\Delta t_{DNS}=0.05$ is used, while at $Re=80$, $\Delta t_{DNS}=0.025$ is used to keep the time step as large as possible while ensuring adequate temporal resolution of the flow dynamics and maintaining numerical stability.

To validate our simulations, we compare our results with reference data available in the literature in Appendix~\ref{sec:codevalidation}.

\subsection{Results}

\subsubsection{Unforced Flow}\label{sec:unforcedflow}
 
Prior to showing the main results of this paper, we shortly discuss the behavior of unforced flow for the chosen test case. The purpose of this is to better understand the flow instability and further motivate our control objective.

\begin{figure*}
\centering
\includegraphics[width=\textwidth, trim= 60pt 300pt 70pt 120pt,clip]{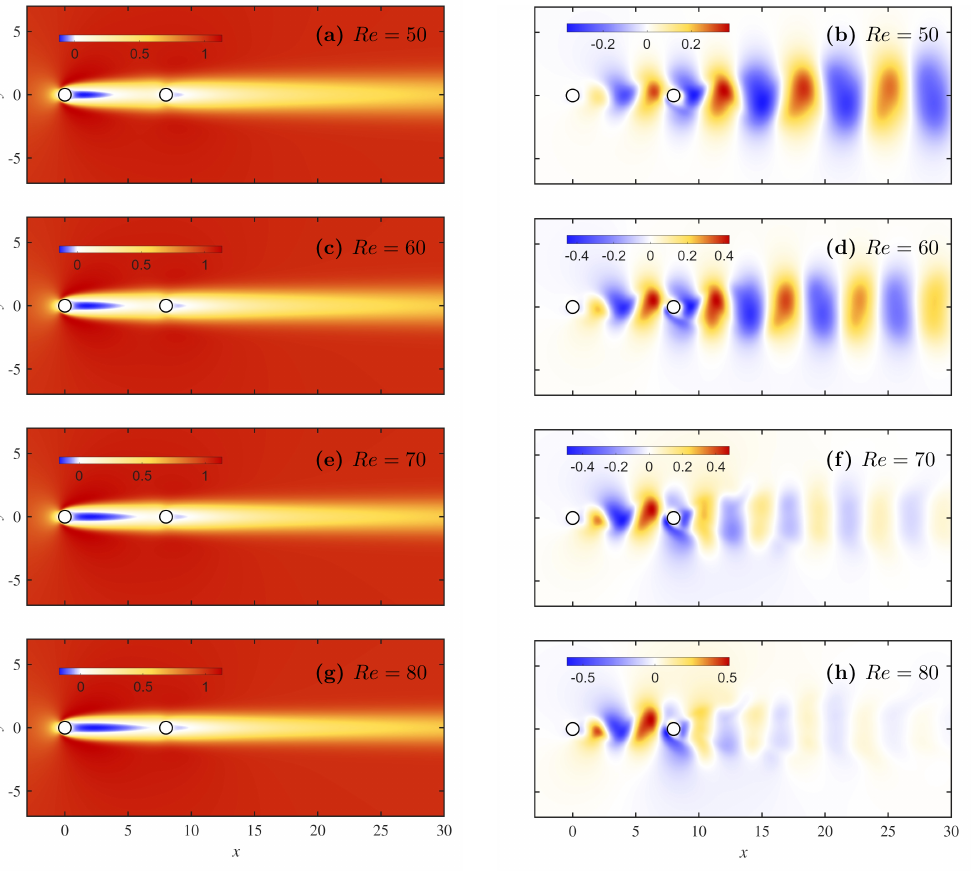}
\caption{ Velocity fields illustrating the steady base flow and the oscillatory flow on the limit cycle at different Reynolds numbers. 
Shown are the $x$-component $u_b$ of the base flow at $Re=50$ (a), $Re=60$ (c), $Re=70$ (e), and $Re=80$ (g), and snapshots of the $y$-component $v'$ of the velocity field oscillating on the limit cycle at $Re=50$ (b), $Re=60$ (d), $Re=70$ (f), and $Re=80$ (h).}
\label{fig:velocity_LC}
\end{figure*}

Figure~\ref{fig:velocity_LC}(a), (c), (e), and (g) shows the base flow for $Re=50, 60, 70$, and~$80$, respectively, which is steady and symmetric. We can see that the region of the base flow with a negative $x$-component $u_b$, which represents the steady recirculation region, increases in length behind both cylinders with increasing $Re$.

The DNS results reveal the unsteady nature of the flow above the critical Reynolds number. Figure~\ref{fig:velocity_LC}(b), (d), (f), and (h) shows instantaneous cross-flow velocity $v'$ at the limit cycle for $Re=50, 60, 70$, and~$80$, respectively. The alternating regions of positive and negative cross-flow velocity $v'$ show the oscillatory nature of the flow, which is physically associated with the von K\'arm\'an vortex street. The instantaneous vorticity of the flow at the limit cycle is shown in Figure~\ref{fig:energy}. From this figure, we observe that vortex shedding occurs both in the gap between the cylinders and in the wake. For the lower Reynolds numbers $Re=50$ and $Re=60$, a standard von K\'arm\'an vortex street is observed behind both the upstream and downstream cylinders. However, at $Re=70$ and $Re=80$, a change in this pattern can be seen. Clockwise and counterclockwise vortex streets about the wake centreline are visible downstream of the second cylinder, with a shear layer developing between them. A similar behavior has been reported by~\citet{wang2010secondary}, who also observed the formation of secondary vortices further downstream, where the two vortex rows merge into a single street and the shedding frequency changes.

This suggests a significant change in the flow structure with increasing Reynolds number, particularly downstream of the second cylinder. This change is also visible in the cross-flow velocity fluctuations $v'$ shown in Figure~\ref{fig:velocity_LC}(b), (d), (f), and (h).

\begin{figure}
\centering
\includegraphics[width=0.7\textwidth, trim=0pt 0pt 0pt 0pt, clip]
{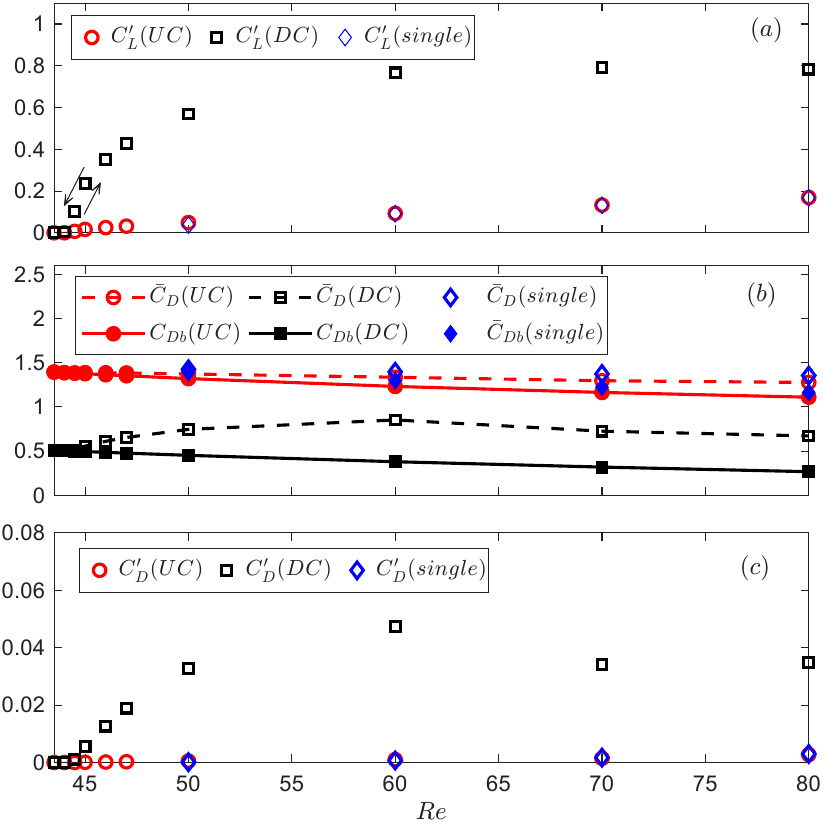}
\caption{Hydrodynamic force coefficients for fully developed limit-cycle flow over two cylinders in tandem configuration with spacing $\gamma=8$, compared with the single-cylinder case, as a function of Reynolds number. 
(a) RMS lift coefficient $C_L'$. 
(b) Mean drag coefficient $\bar{C}_D$ and steady drag coefficient $C_{Db}$. 
(c) RMS drag coefficient $C_D'$. }
\label{fig:Loads_unforced}
\end{figure}

Figure~\ref{fig:Loads_unforced} shows the hydrodynamic force coefficients acting on the two cylinders at the limit cycle versus the Reynolds number. These forces are decomposed into mean and fluctuating components. Here we denote the mean lift and drag coefficients by $\bar{C}_L$ and $\bar{C}_D$, respectively, and the root mean square of their fluctuating components by $C_L'$ and $C_D'$. 

Note that the mean lift forces $\bar{C}_L$ acting on both cylinders are zero due to the symmetry of the mean flow. However, the vortex shedding induces fluctuating forces on both cylinders. 
The RMS of the lift and drag coefficients, $C_L'$ and $C_D'$, acting on the upstream cylinder are similar to those in the single-cylinder case. However, the fluctuating loads on the downstream cylinder are much higher, as shown in Figures~\ref{fig:Loads_unforced}(a) and (c), since it is immersed in the unsteady wake of the upstream cylinder.  

The unsteadiness of the flow also induces higher mean drag coefficients compared to the corresponding base-flow drag $C_{Db}$ for both cylinders at all Reynolds numbers (see Figure~\ref{fig:Loads_unforced}(b)). 
However, the mean drag coefficient $\bar{C}_D$ is lower on the downstream cylinder because it is located in the wake of the upstream cylinder, where the mean velocity is lower than the freestream velocity.

Overall, the unsteadiness of the flow generates significant fluctuating loads and increases the mean drag on the cylinders, with particularly strong unsteady loading on the downstream cylinder due to the tandem-cylinder configuration. Therefore, suppressing vortex shedding and stabilizing the flow both in the gap and in the wake is of great interest, as it can minimize its adverse effects.

In a supercritical Hopf bifurcation, oscillations emerge as the Reynolds number exceeds the critical value $Re_c$, and the flow subsequently settles onto a limit cycle whose amplitude slowly increases with increasing Reynolds number. Consistent with this behavior, $C_l'$ increases slowly with $Re$ in the vicinity of the critical point.
To verify the nature of the bifurcation, direct numerical simulations (DNS) were performed for $Re=43.5$–$45$ with increments of $0.05$. The Reynolds number was first increased from $Re=43.5$ to $Re=45$, and then decreased over the same range to check for possible hysteresis in $C_l'$, as indicated by the arrows in the Figure~\ref{fig:Loads_unforced}(a). A nonzero value of $C_l'$ was first observed at $Re=44.5$, and no hysteretic behavior was detected during the decrease of $Re$, which confirms that the bifurcation is supercritical.

\subsection{Reduced-Order Model}
Here we present the construction of the reduced-order model of the flow using the weakly nonlinear analysis from Section~\ref{sec:WNA} and compare it to the DNS results. 

Starting from an initial guess around $Re \approx 44.5$, where the first instability is observed in DNS (see Figure~\ref{fig:Loads_unforced}), we determine the critical Reynolds number $Re_c = 44.1$, consistent with the value reported in the literature~\cite{wang2022first}, by iteratively solving the first two orders of the weakly nonlinear analysis. The corresponding eigenvalue is $\lambda_0=-0.0004+0.6756\imath$.

\begin{figure}
\centering
\includegraphics[width=0.9\textwidth, trim=60pt 390pt 50pt 70pt, clip]{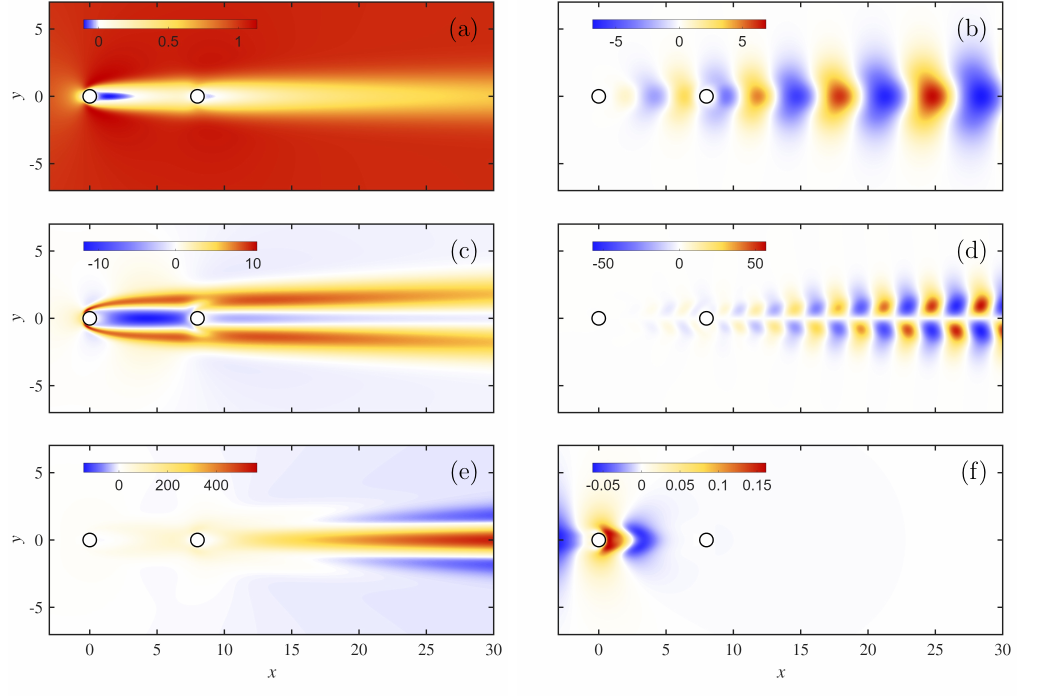}
\caption{Spatial structure of the modes appearing in the expansion~\eqref{eq:solution_wna1} up to the order $\epsilon$.
(a) The $x$-component $u_0$ of the base flow at $Re_c$
(b) Real part $\Re (v_1^{A})$ of the $y$-component of the critical global mode
(c) The $x$-component $u_2^1$ of the base flow modification
(d) Real part $\Re (v_1^{A^2})$ of the $y$-component of the second harmonic mode, 
(e) The $x$-component $u^{|A|^2}$ of the shift mode
(f) Real part $\Re (v_1^{A \ast})$ of the $y$-component of the critical adjoint global mode}
\label{fig:modes}
\end{figure}

The modes up to order $\epsilon$ in the weakly nonlinear expansion~\eqref{eq:solution_wna1} of the flow are shown in Figure~\ref{fig:modes}. The $x$-component $u_0$ of the base flow $\mathbf{U}_0$ at $\epsilon=0$ (or $Re_c=44.1$) is displayed in Figure~\ref{fig:modes}(a).
 The $x$-component $u_2^1$ of the modification of $\mathbf{U}_0$ corresponding to the unstable equilibrium at $\epsilon>0$ is shown in Figure~\ref{fig:modes}(c). The negative values of $u_2^1$ in the gap and the wake indicate that the spatial extent of the recirculation region increases with $\epsilon$ (or $Re$), consistent with the behavior observed in Figure~\ref{fig:velocity_LC}(a).
 
Figure~\ref{fig:modes}(b) shows the $y$-component (real part) $\Re(v_1^A)$ of the critical global mode $\mathbf{U}_1^A$. This mode oscillates approximately at the fundamental frequency of the flow and captures the large-scale structures of its unsteady component. The second harmonic mode $\mathbf{U}_2^{A^2}$, which oscillates at twice the fundamental frequency, is characterized by smaller spatial structures, as illustrated in Figure~\ref{fig:modes}(d), which shows its $y$-component (real part) $\Re(v_2^{A^2})$.

 The $x$-component $u_1^{|A|^2}$ of the shift mode, which approximates the difference between the mean flow and the base flow, is plotted in Figure~\ref{fig:modes}(e).
The positive values of $u_1^{|A|^2}$ in the gap and the wake indicate a decrease in the recirculation bubble length in the mean flow compared to the base flow, consistent with observations for a single cylinder in~\citet{sipp2007global}.

Finally, Figure~\ref{fig:modes}(f) shows the $y$-component $\Re(v_1^{A\star})$ of the critical adjoint global mode at $Re_c$. Its largest values are concentrated near the upstream cylinder, indicating that this region is the most effective location for applying volume forcing for flow control.

The modal structure exhibits a clear symmetry with respect to the centerline $y=0$: modes arising at odd orders in the expansion are antisymmetric, while those at even orders are symmetric. This behavior is consistent with the weakly nonlinear analysis for the single-cylinder wake reported by~\citet{sipp2007global}, and is also evident from the modal structures shown in Figure~\ref{fig:modes}.

As a consequence, the $\mathcal{O}(\epsilon)$ contributions, which are symmetric, vanish when projected onto the adjoint global mode $\mathbf{u}_1^{A\star}$, which is antisymmetric. This explains the cancellation stated in Section~\ref{sec:estimation} and yields the explicit expression for $\tilde{A}$ in~\eqref{eq:Atilde}.

The coefficients of the Stuart–Landau model~\eqref{eq:SL_resonance_t} representing this flow are listed in Table~\ref{tab:SLcoeff}. They are computed from~\eqref{eq:SLcoefficients} in Appendix~\ref{sec:WNAappendix}. The positive value of $\Re(a_0)$ indicates that the fixed point $A=0$ is repelling for $\epsilon>0$, which is consistent with the DNS observation that the flow undergoes a supercritical Hopf bifurcation. The critical global mode is normalized such that $v_1^A(0.65,0)=-0.0979+0.2661\imath$, yielding $\Re(a_1)\approx\Re(a_0)$ and therefore a magnitude of limit-cycle amplitude close to unity (see expression~\eqref{eq:LC_magnitude}). For the chosen mesh, $|A|_{LC}=1.0374$. The coefficient $a_2$ is computed for the forcing structure discussed in Section~\ref{sec:optimalforcing}.

For the unforced flow, we compare the complex amplitude $A$ predicted by the reduced model with the amplitude $\tilde{A}$ evaluated from DNS using~\eqref{eq:Atilde}, assuming full knowledge of the velocity field over the entire computational domain. The difference 
\[
e_A \equiv \tilde{A}-A,
\]
between these two quantities depends on the approximation error $\mathbf{e}_{\mathbf{u}}$ in~\eqref{eq:approximationerror}. In particular,
\begin{align}\label{eq:DNS_projected_resonant}
e_A := \frac{1}{\sqrt{\epsilon}} 
\langle \mathbf{u}_1^{A \star}, \mathbf{e}_{\mathbf{u}} \rangle 
e^{-\imath \omega_0 t}.
\end{align}

The DNS is initialized close to the base flow with the perturbation velocity $
\mathbf{u}'(t=0)=0.0015\,\mathbf{u}_1^A$,
which lies in the subspace of the critical global mode. This initialization is used for all Reynolds numbers and corresponds to $\sqrt{\epsilon}A=0.0075$. Consequently, the approximation error $\mathbf{e}_{\mathbf{u}}(t=0)$ in~\eqref{eq:approximationerror} is initially zero.

The evolution of $|\tilde{A}|$ for $Re=50$, $60$, $70$, and $80$ is shown in Figures~\ref{fig:control_full}(a)--(d). The model captures the overall trend of the amplitude dynamics and reproduces the qualitative approach to the limit cycle. 

\begin{figure}
\centering
\includegraphics[width=0.7\textwidth,trim=0pt 0pt 0pt 0pt,clip]{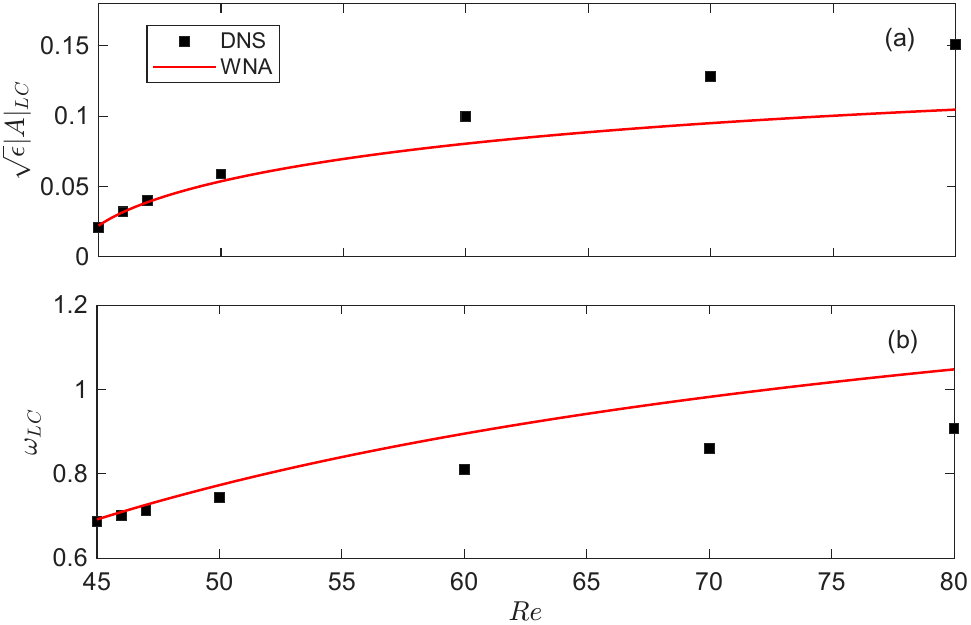}
\caption{Prediction of the (scaled) 
mode amplitude $\sqrt{\epsilon}|A|_{LC}$ and the shedding frequency $\omega_{LC}$ at the natural limit cycle vs. $Re$. The values of $A_{LC}$ are estimated as $\tilde{A}_{LC}$ from DNS assuming complete knowledge of the velocity field within the entire computational domain, and obtained from the Stuart-Landau model with coefficients derived from our weakly nonlinear analysis (WNA).}
\label{fig:DNSvsWNA}
\end{figure}

The unforced Stuart--Landau model, however, underpredicts the magnitude of the limit-cycle amplitude $|A|_{LC}$. Moreover, the discrepancy between the model and DNS increases with the Reynolds number, as shown in Figure~\ref{fig:DNSvsWNA}, which displays the variation of $\sqrt{\epsilon}|A|_{LC}$ and $\omega_{LC}$ with $Re$. The main source of this increasing modeling error is the unsteady flow distortion, evident in the vorticity fields in Figures~\ref{fig:energy}(c)-(d), which is not captured accurately by the weakly nonlinear approximation. In particular, for $Re \gtrsim 70$ the flow deviates significantly from its first-order approximation based on the global mode at $Re_c$, as discussed in Section~\ref{sec:unforcedflow}.

\begin{table}
\centering
 \caption{Values of the coefficients of the Stuart-Landau model obtained via weakly nonlinear analysis (WNA). The coefficient $a_2$ is computed for the localized forcing structure~\eqref{eq:localforcing}, where $\Omega_s$ is the union of circular domains $\Omega_{s_\pm}$ of radius $0.07$ centered at $\mathbf{x}_i=(0.32,\pm 0.59)$ (Figure~\ref{fig:a2norm}).} \label{tab:SLcoeff}

\begin{tabular}{lcccc}
$Re_c$ & $\omega_0$ & $a_0$   &  $a_1$  &  $a_2$\\ [5pt] \hline
 $44.1$ & $0.6756$& $7.6571+2.9857\imath$ & $7.1153-31.2244\imath$ & $0.0368-0.0645\imath$ 
\end{tabular}
\end{table}

\subsection{Closed-Loop Control}

Here, we apply the control design introduced in Section~\ref{sec:control} to suppress vortex shedding in both the gap flow and the wake of the two cylinders. We consider the same four Reynolds numbers as in the previous section, namely, $Re=50, 60, 70$, and $80$.

\begin{figure}
\centering
\includegraphics[width=0.7\textwidth,trim=45pt 95pt 20pt 50pt,clip]{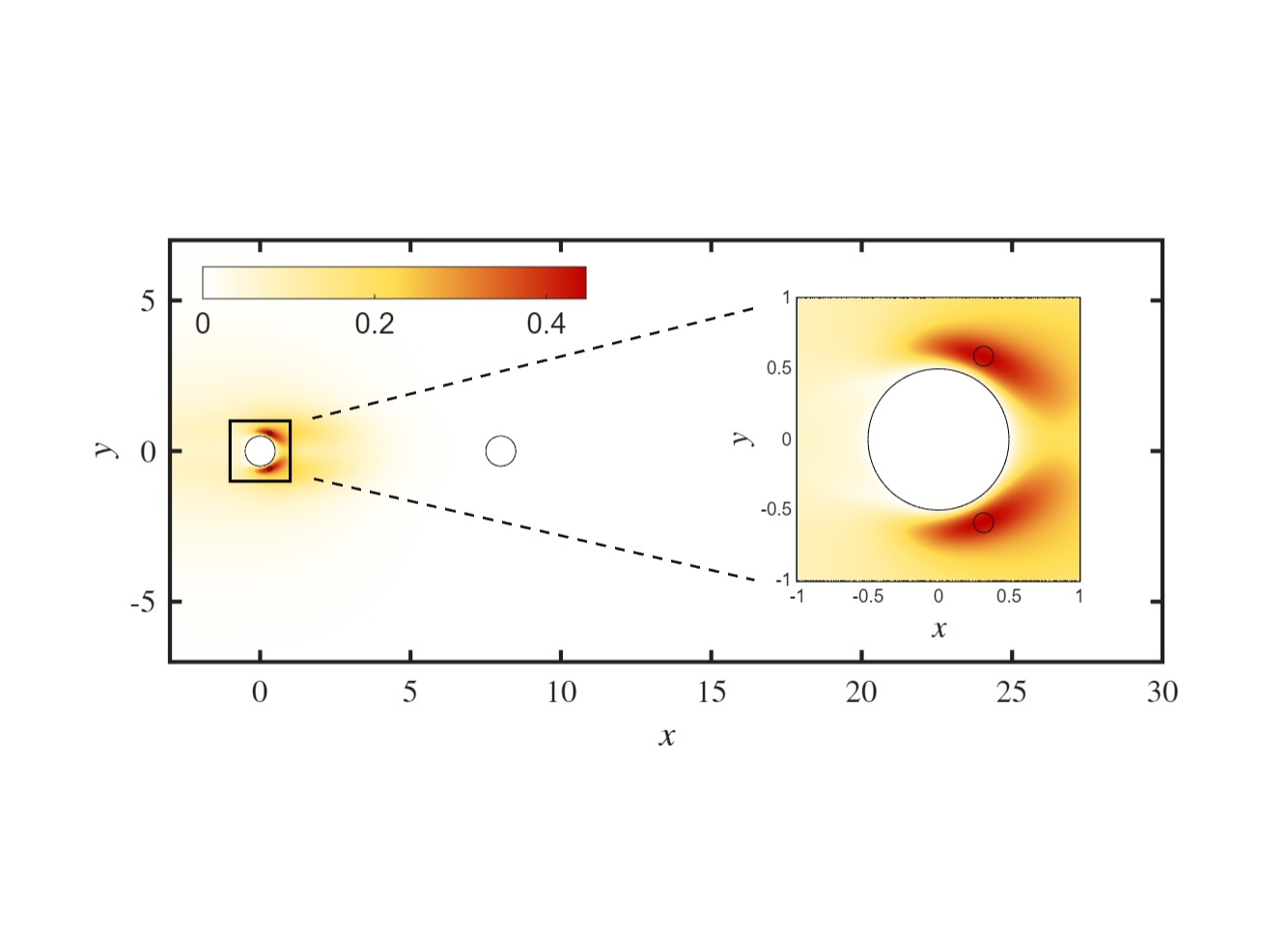}
\caption{Spatial distribution of the norm of the adjoint mode $||\mathbf{u}_1^{A \star}(\mathbf{x})||$. Circles indicate our choice of localized forcing domain $\Omega_s$, with chosen radius $0.07$ and centered on locations $\mathbf{x}_i=(0.32, \pm 0.59)$ where this norm is the highest, thus representing favorable locations for forcing.}
\label{fig:a2norm}
\end{figure}

To ensure physical realizability, we employ a spatially-localized forcing structure rather than the distributed optimal volume forcing. As shown in Section~\ref{sec:optimalforcing}, maximizing the control effectiveness requires choosing a location $\mathbf{x}_i$ where the norm of the adjoint mode,  $||\mathbf{u}_1^{A\star}(\mathbf{x}_i)||$, is large. From Figure~\ref{fig:a2norm}, this location is approximately $\mathbf{x}_i=(x_i,y_i)=(0.32,\pm 0.59)$. We therefore employ the localized forcing defined in \eqref{eq:localforcing}, with $\Omega_s$ taken as the union of the circular domains $\Omega_{s_\pm}$ of radius $0.07$ centered at $\mathbf{x}_i=(0.32,\pm 0.59)$, indicated by the circles in Figure~\ref{fig:a2norm}.

Before applying any control action, we run the unforced DNS for $500$ time units for each Reynolds number to ensure that the flow is fully settled on the limit cycle. The controller is then activated at $t=500$, starting from this limit-cycle state.
 
 The time-discretized sequence of control inputs $\mathbf{q}_j=[\Re(E_j), \Im(E_j)]^T$ is determined by the model predictive control scheme described in Section~\ref{sec:mpc}, which minimizes the cost function~\eqref{eq:costfunction}. The control actions are applied to the continuous-time system using a zero-order hold. To solve the minimization problem, predictions of the future states $\mathbf{X}_{j+k}=[\Re(A_{j+k}), \Im(A_{j+k})]^T$ are obtained by numerically integrating the Stuart--Landau model~\eqref{eq:SL_resonance_t}.

Following our methodology, the feedback controller is constructed on the slow time scale. Accordingly, the MPC operates with a larger time step than the forced DNS, and we set $\Delta t_{\mathrm{MPC}} = 20\,\Delta t_{\mathrm{DNS}}$. Furthermore, the weighting matrix $\mathbf{R}_{\Delta u}$ is used to limit the temporal gradients of $E$. For $Re=50$ and $Re=60$, we use $\Delta t_{\mathrm{DNS}}=0.05$, as in the unforced simulations, and $\mathbf{R}_{\Delta u}=0.1$. For $Re=70$ and $Re=80$, where wake suppression becomes more challenging, the smaller DNS time steps $\Delta t_{\mathrm{DNS}}=0.025$ and $\Delta t_{\mathrm{DNS}}=0.02$ are used together with a reduced control weight $\mathbf{R}_{\Delta u}=0.08$ to facilitate suppression.

Regarding the remaining weighting matrices, we set $\mathbf{Q}=1000\mathbf{I}$ for all Reynolds numbers. The control weight is chosen as $\mathbf{R}=0.0015\mathbf{I}$ for $Re=50$ and $Re=60$, and $\mathbf{R}=0.0008\mathbf{I}$ for $Re=70$ and $Re=80$, allowing higher forcing amplitudes for higher Reynolds numbers. The prediction horizon is $m=20$ time steps.

In the following, we consider two measurement settings: (i) full-field velocity measurements and (ii) pointwise velocity  measurements at a finite number of sensor locations. All controller parameters described above are kept identical in both cases.

\subsubsection{Control with full measurements}

We first assume the ideal case of full-domain measurements, i.e.\ complete knowledge of the velocity field throughout the computational domain. The reduced state $\tilde{\mathbf{X}}_j=[\Re(\tilde{A}_j), \Im(\tilde{A}_j)]^T$ used in the feedback law is then evaluated from the velocity field of the forced DNS solution using~\eqref{eq:Atilde}. This provides a benchmark for the best achievable performance of the control strategy.

\begin{figure*}
\centering
\includegraphics[width=1\textwidth,trim= 0pt 20pt 0pt 0pt, clip]{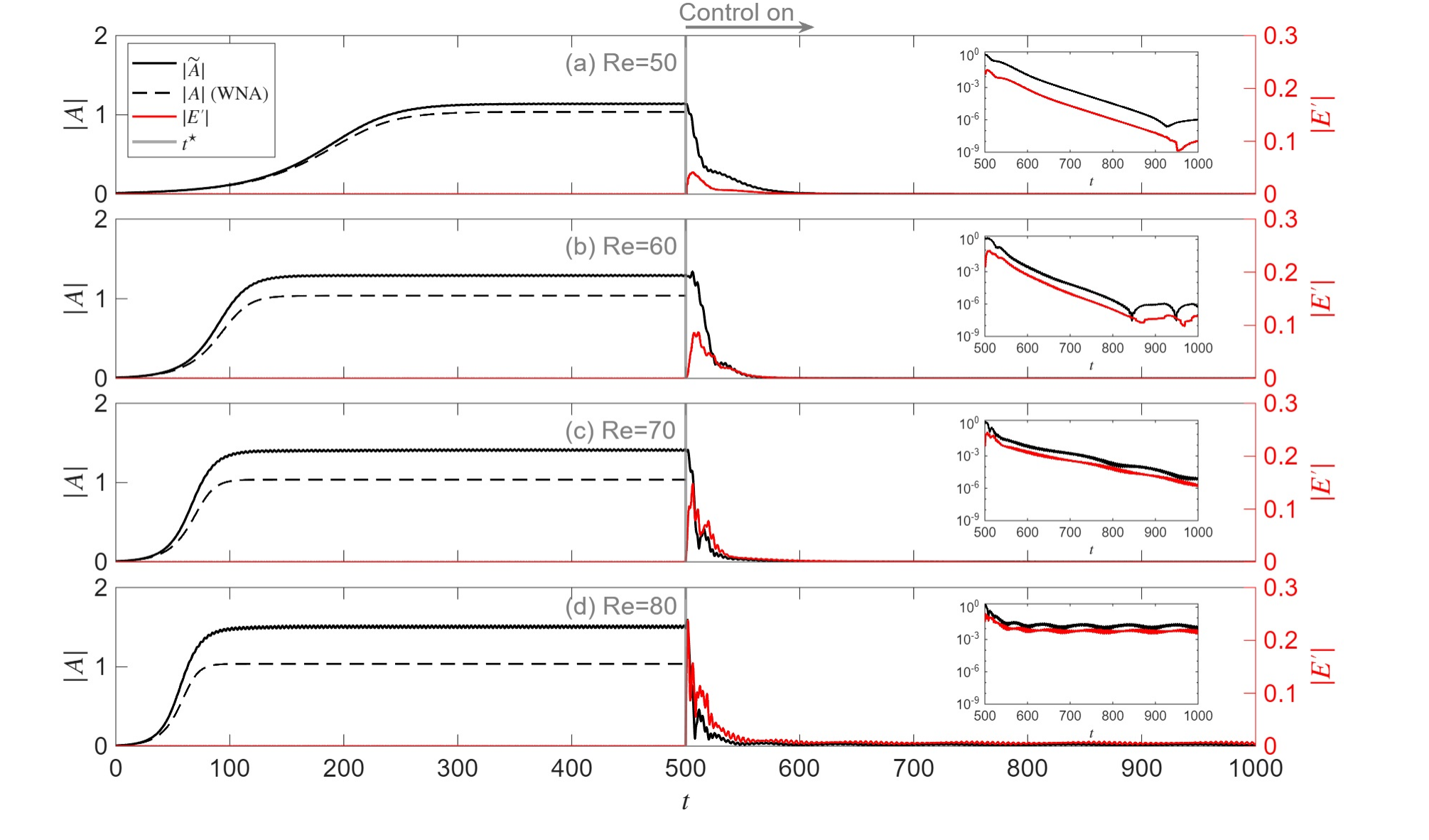}
\caption{Time evolution of the global mode amplitude and its suppression. For $t\leq 500$, the unforced evolution is shown, comparing the amplitude obtained from DNS, the estimate $\tilde{A}$ from full velocity measurements~\eqref{eq:Atilde}, and the prediction from the weakly nonlinear analysis (WNA). At $t=500$, control is activated, and the subsequent decay of the amplitude is achieved using MPC with resonant forcing of amplitude $E'$ and feedback based on $\tilde{A}$. (a) $Re=50$, (b) $Re=60$, (c) $Re=70$, (d) $Re=80$.}
\label{fig:control_full}
\end{figure*}

We track the time evolution of the reduced state and reduced input with control applied over the interval $t=500$ to $t=1000$. Figure~\ref{fig:control_full} shows their magnitudes, $|\tilde{A}|$ and $|E'|$. Here, we choose to plot forcing amplitude $E'=\epsilon^{3/2}E$ so that the control amplitudes are directly comparable across Reynolds numbers.

This time interval is sufficient to reduce $|\tilde{A}|$ to very small values, of order $10^{-7}$ for $Re=50$ and $Re=60$, and $10^{-6}$ for $Re=70$. The corresponding magnitudes of the forcing amplitude $E'$ are of order $10^{-8}$ for $Re=50$ and $Re=60$, and $10^{-6}$ for $Re=70$.

For $Re=80$, the system is not stabilized to the same extent; instead, it settles into another oscillatory state with a mean $|\tilde{A}|$ of order $10^{-2}$, i.e.\ about two orders of magnitude smaller than the unforced limit-cycle amplitude ($|\tilde{A}|_{LC}\approx 1.5$). The corresponding amplitude of the reduced input is of the same order.

\begin{figure*}
\centering
\includegraphics[width=\textwidth, trim = 20pt 50pt 0pt 100pt, clip=true]{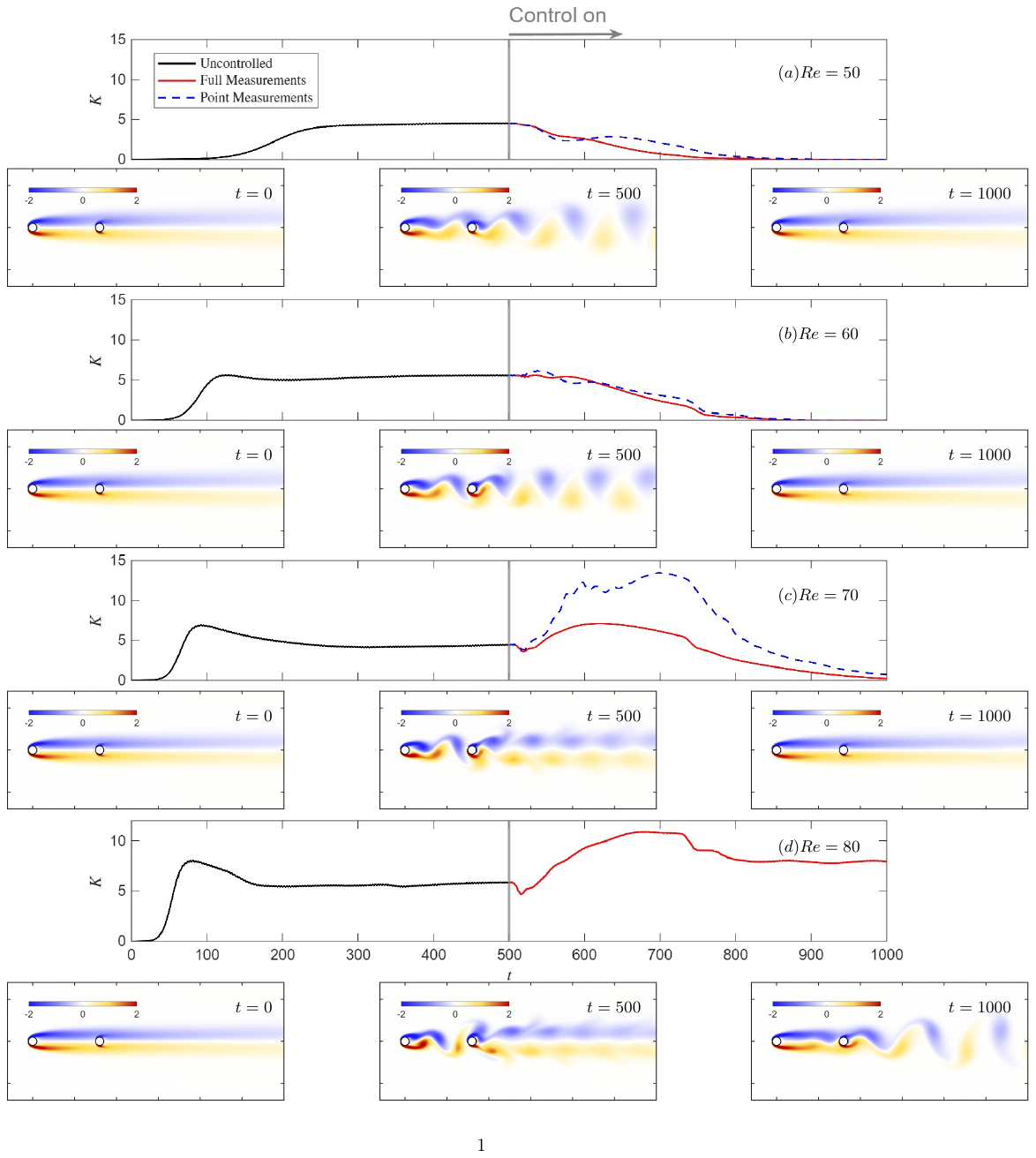}
\caption{Evolution of energy $K=0.5 \langle \mathbf{u}', \mathbf{u}' \rangle $ of the unsteady velocity $\mathbf{u}'$ relative to the base flow $\mathbf{u}_b$, as $\mathbf{u}'=\mathbf{u}-\mathbf{u}_b$. The flow evolves to the limit cycle before $t=500$, at which time control is turned on. The solid red curve represents results  obtained by applying control  using the full measurements for the feedback, while the dashed blue curve is the result with state estimation via point measurements only. Instantaneous vorticity and streamlines for times $t=0$, $t=500$ (when the flow is at the limit cycle), and $t=1000$ are shown. 
(a) $Re=50$ (b) $Re=60$ (c) $Re=70$ (d) $Re=80$.}
\label{fig:energy}
\end{figure*}

We examine the fluctuation energy 
\[
K = \tfrac{1}{2}\langle \mathbf{u}', \mathbf{u}' \rangle
\]
of the unsteady velocity $\mathbf{u}' = \mathbf{u} - \mathbf{u}_b$. 
By definition, $K=0$ when the flow is at its steady equilibrium. Figure~\ref{fig:energy} shows the evolution of $K$ for the unforced flow up to $t=500$ and for the controlled flow over the interval $t=500$ to $t=1000$. 

For $Re=50$, $60$, and $70$, the energy at $t=1000$ is significantly reduced, reaching values of order $10^{-8}$, $10^{-7}$, and $10^{-1}$, respectively. For $Re=70$, the energy continues to decrease under prolonged control (not shown here), reaching values of order $10^{-2}$ after an additional $100$ time units. In contrast, for $Re=80$, a substantial level of energy remains in the flow.

To further visualize the controlled dynamics, we plot the instantaneous vorticity at $t=0$, $t=500$ (corresponding to the limit cycle), and $t=1000$. At $t=0$, the vorticity of the base flow is shown; the actual flow differs only by a small perturbation, which is not visually discernible.

For $Re=50$, $60$, and $70$, the control drives the flow close to the steady equilibrium, with no visible difference between vorticity plots at $t=0$ and $t=1000$. For $Re=80$, however, the flow settles into another periodic state, as also observed in Figure~\ref{fig:control_full}(d), with persistent vortex shedding. Compared to the natural limit cycle at $t=500$, the vortex formation length is increased and the shedding in the gap region is significantly weakened.

\begin{figure}
\centering
\includegraphics[width=0.6\textwidth]{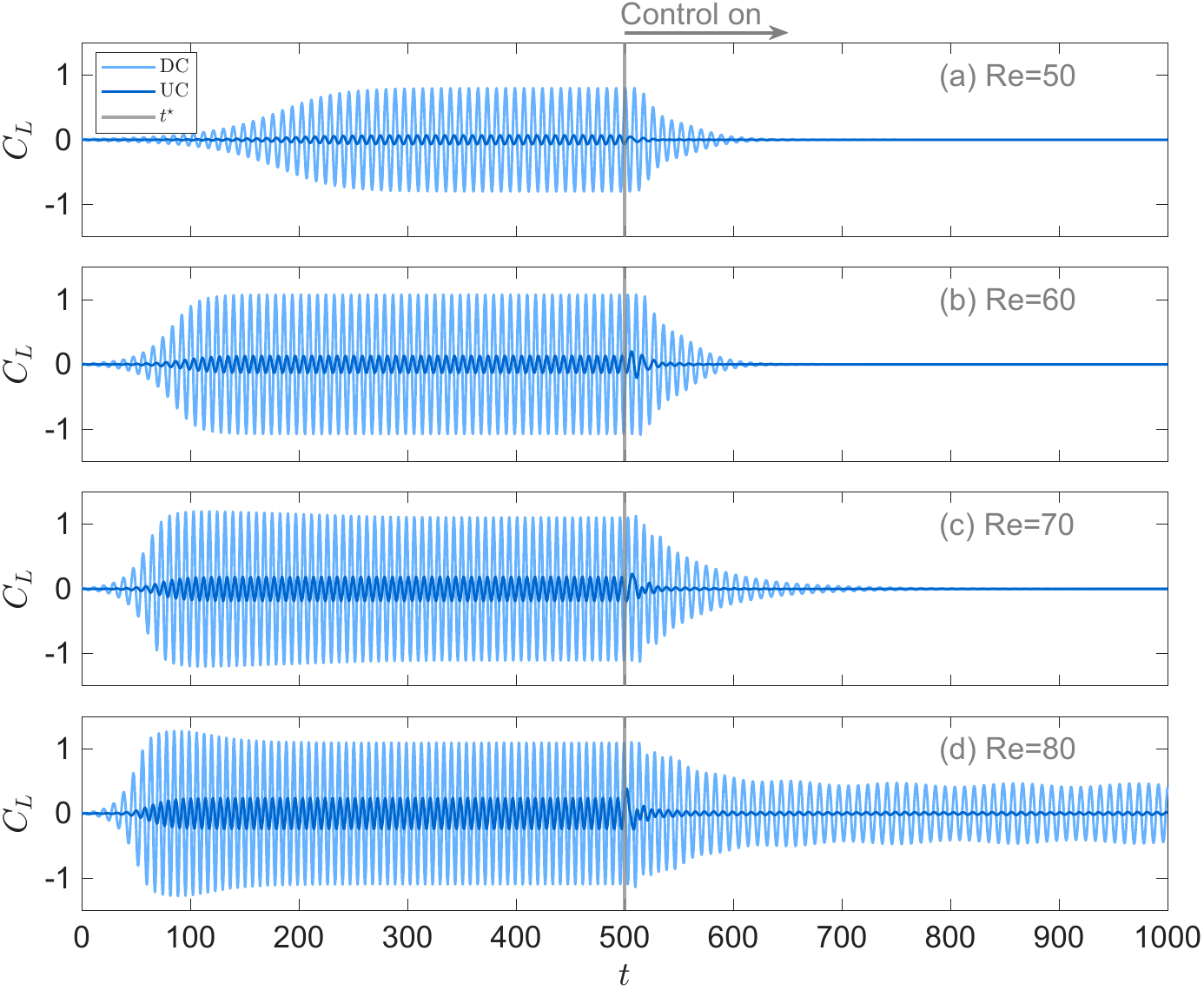}
\caption{Time evolution and suppression of lift coefficient $C_L$ acting on the upstream (UC) and downstream (DC) cylinders. The MPC feedback for the MPC is based on full velocity measurements.(a) $Re=50$ (b) $Re=60$ (c) $Re=70$, (d) $Re=80$. }
\label{fig:CL_full}
\end{figure}

\begin{figure}
\centering
\includegraphics[width=0.6\textwidth]{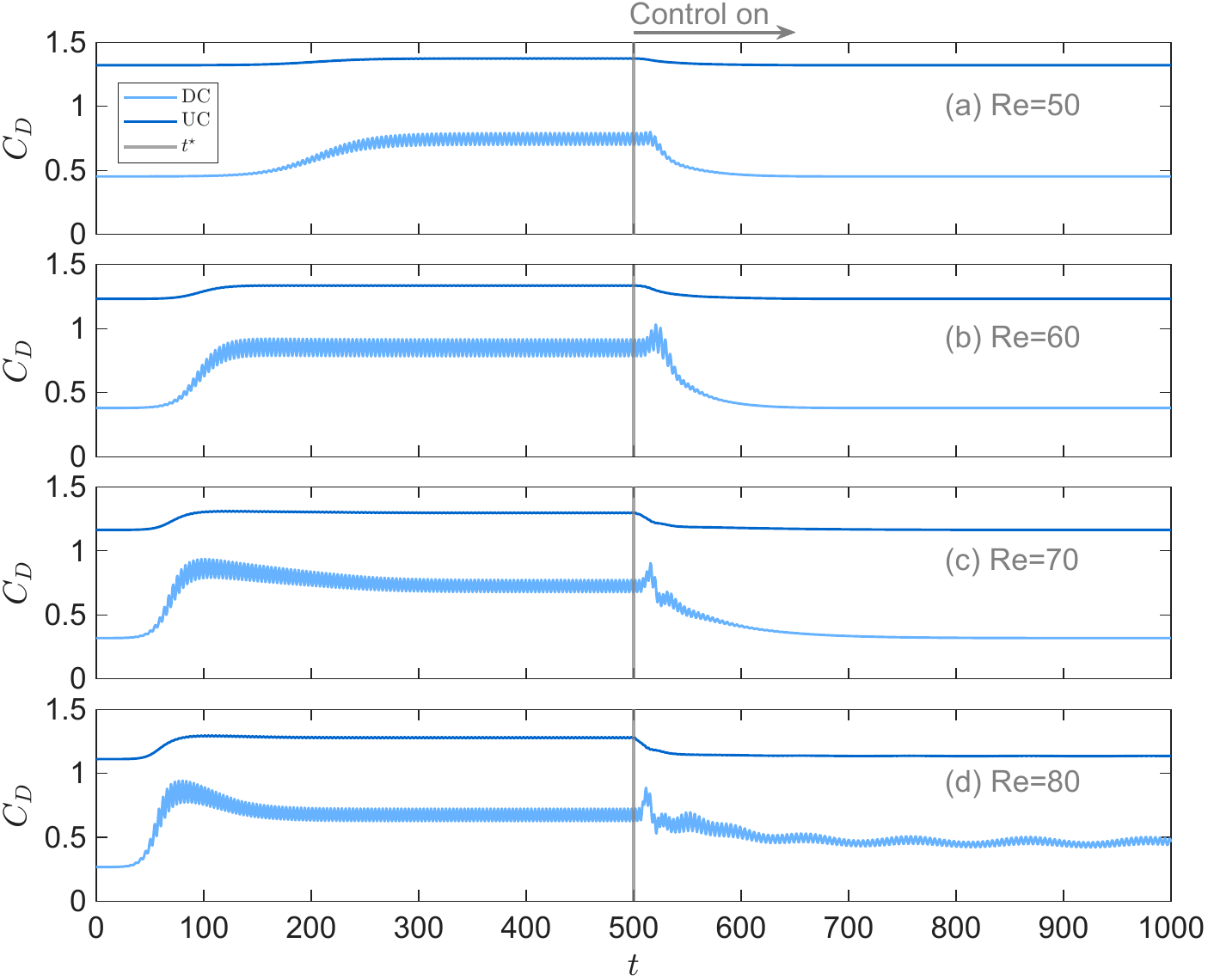}
\caption{Time evolution and suppression of drag coefficient $C_D$ acting on the upstream (UC) and downstream (DC) cylinders. The MPC feedback for the MPC is based on full velocity measurements.(a) $Re=50$ (b) $Re=60$ (c) $Re=70$, (d) $Re=80$. }
\label{fig:CD_full}
\end{figure}
Finally, we assess the impact of control on the lift and drag coefficients. Figures~\ref{fig:CL_full} and~\ref{fig:CD_full} show the time evolution of $C_L$ and $C_D$ for both the unforced and controlled flows. Consistent with the vorticity observations, for $Re=50$, $60$, and $70$, the lift coefficient is reduced to near zero for both cylinders, reaching values of less than $10^{-7}$ for $Re=50$, less than $10^{-6}$ for $Re=60$, and less than $10^{-4}$ for $Re=70$, while the drag coefficient approaches its steady value.

For $Re=80$, the reduction in the lift coefficient is noticeably less pronounced; however, it is still significantly mitigated compared to the unforced case. For the upstream cylinder, the RMS of lift coefficient is reduced from $C_L' \approx 0.17$ in the unforced limit cycle to $C_L' \approx 0.014$ under control. For the downstream cylinder,  is $C_L'$ reduced by more than a factor of two, from $C_L' \approx 0.78$ in the unforced case to $C_L' \approx 0.31$.

The drag coefficient of the upstream cylinder closely approaches its steady value ($\bar{C}_D \approx 1.11$), with a controlled mean of approximately $\bar{C}_D \approx 1.13$ (compared to $\bar{C}_D \approx 1.27$ in the unforced case) and small residual oscillations of order $10^{-3}$ ($C_D' \approx 9\times 10^{-4}$ compared to $2.7\times 10^{-3}$ in the unforced case).For the downstream cylinder, the mean drag is reduced from $\bar{C}_D \approx 0.67$ in the unforced case to $\bar{C}_D \approx 0.45$, approaching an intermediate value between the steady-state value ($\bar{C}_D \approx 0.26$) and the unforced limit cycle. The corresponding drag fluctuations are also reduced ($C_D' \approx 0.02$ compared to $0.035$ unforced), although noticeable oscillations persist.

\subsubsection{Control with point measurements}

In the previous section, we used the reduced state estimated from the DNS velocity field to design the feedback law. However, measuring the velocity field over a large domain, for example using particle image velocimetry (PIV), is often too slow for real-time control. In practical applications, the flow is instead measured using a limited number of spatially localized sensors, which provide only partial information about the velocity field. We therefore consider the use of point measurements of the velocity.

To this end, we seek suitable sensor locations that enable an accurate estimation of the reduced state. The goal is to compute $\tilde{A}_p$ via the mapping~\eqref{eq:measurements}, such that it closely approximates $\tilde{A}$ obtained from full-domain measurements using~\eqref{eq:Atilde}, while using as few point measurements $u_1^A(\mathbf{x}_i)$ and/or $v_1^A(\mathbf{x}_i)$ as possible. In this work, we restrict ourselves to the minimum number of measurements required by the estimation approach in~\eqref{eq:measurements}, namely two.

In analogy with~\eqref{eq:approximationerror}, we define the approximation error
\begin{equation}\label{eq:error}
\tilde{\mathbf{e}}\equiv(\tilde{\mathbf{e}}_{\mathbf{u}}, \tilde{e}_p):=\mathbf U-\mathscr{G}(\tilde{\mathbf{X}}),
\end{equation}
where $\mathscr{G}$ is given by~\eqref{eq:Gmap}. For $\tilde{A}_p$ to closely approximate $\tilde{A}$, it is required that $\mathbf{G}^{\dagger}\tilde{\mathbf{e}}_{\mathbf{u}}$ remains small, where $\mathbf{G}$ is defined in~\eqref{eq:measurements}.

Taking this criterion into account, we select the following measurement configurations: \\$(u_1^A(3.8,1.4),\, v_1^A(3.8,1.4))$ for $Re=50$, $(v_1^A(2.8,0),\, v_1^A(1.4,0))$ for $Re=60$, and $(v_1^A(2.8,0),\, v_1^A(1.2,0))$ for $Re=70$. For $Re=50$, we use both velocity components at a single point, which is advantageous from an application standpoint since it requires only one sensor location. However, estimating $A$ from a single point is more challenging, and at higher Reynolds numbers this approach does not provide sufficient accuracy. Therefore, for $Re=60$ and $Re=70$, we instead use measurements of the $y$-velocity component at two distinct locations, which leads to smaller values of $\mathbf{G}^{\dagger}\tilde{\mathbf{e}}_{\mathbf{u}}$ and thus a more accurate estimation of~$A$.

At the selected measurement locations, the higher-order modes $\mathbf{U}_2^{|A|^2}$ and $\mathbf{U}_2^{A^2}$ are small, as shown in Figure~\ref{fig:modes}. This justifies the use of the linear mapping~\eqref{eq:measurements}, which retains only terms proportional to $A$, for the state estimation.

Control with point measurements is not considered for $Re=80$, due to the large state-estimation error arising from the altered structure of the global mode at this Reynolds number, which is not adequately captured by the present model.

\begin{figure*}
\centering
\includegraphics[width=\textwidth,trim= 0pt 130pt 0pt 0pt, clip]{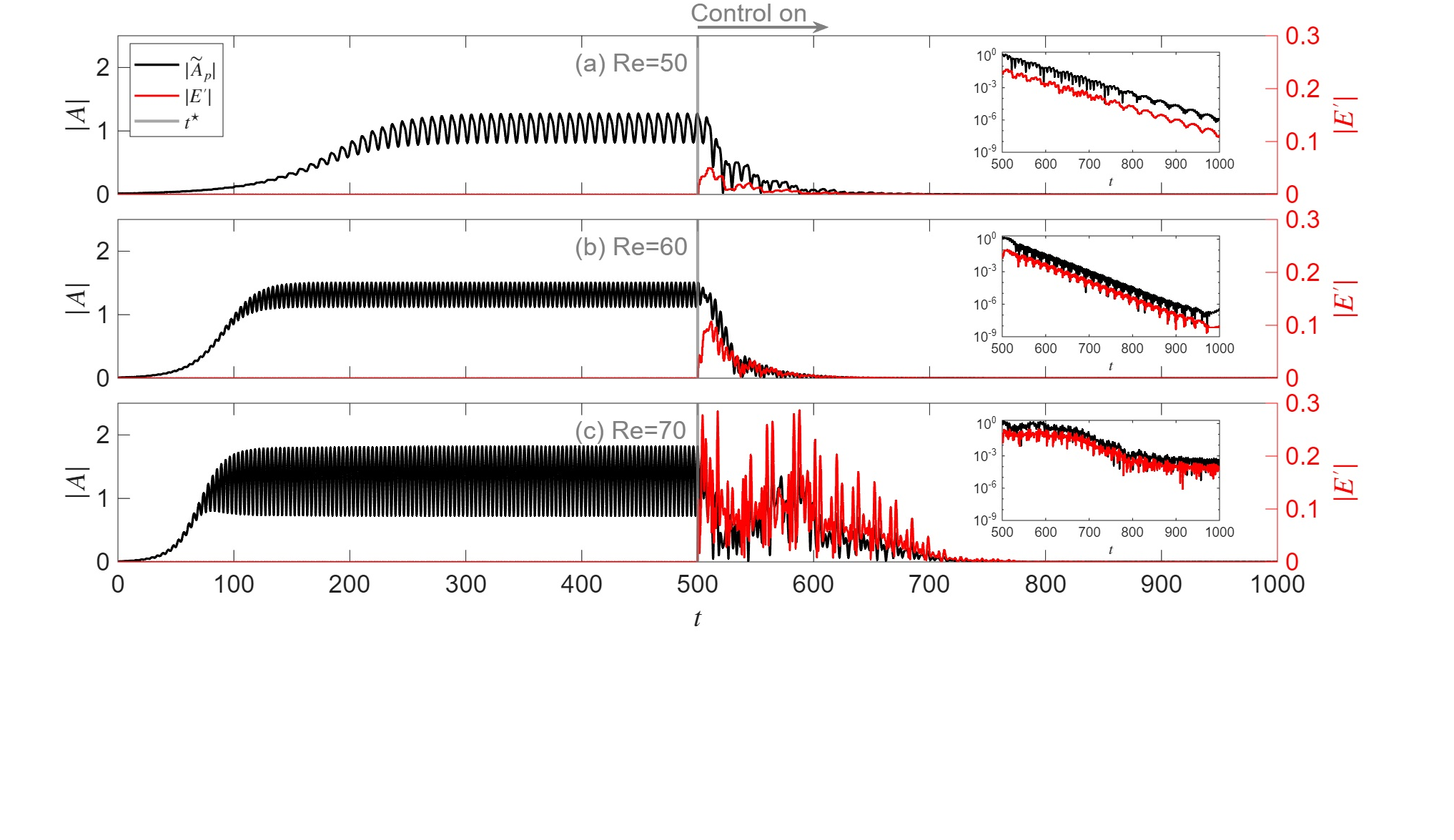}
\caption{Time evolution of the global mode amplitude and its suppression. For $t\leq 500$, the estimate $\tilde{A}_p$ obtained from pointwise velocity measurements~\eqref{eq:measurements} in the unforced flow is shown. The following measurement configurations are used: $(u_1^A(3.8,1.4),\, v_1^A(3.8,1.4))$ for $Re=50$, $(v_1^A(2.8,0),\, v_1^A(1.4,0))$ for $Re=60$, and $(v_1^A(2.8,0),\, v_1^A(1.2,0))$ for $Re=70$. At $t=500$, control is activated, and the subsequent decay of the amplitude is achieved using MPC with resonant forcing of amplitude $E'$ and feedback based on $\tilde{A}_p$. (a) $Re=50$, (b) $Re=60$, (c) $Re=70$.}
\label{fig:control_meas}
\end{figure*}

Figure~\ref{fig:control_meas} shows the time evolution of the amplitude of the reduced state $|\tilde{A}_p|$, estimated from the selected point measurements, for the unforced flow over the interval $t=0$--$500$. An oscillatory discrepancy, compared to the amplitude obtained from full measurements (Figure~\ref{fig:control_full}), is observed in $\tilde{A}_p$ and grows as the flow approaches the limit cycle. This behavior does not reflect the physical flow dynamics, but arises from the increasing approximation error in the state estimation.

The estimation error is significantly larger at $Re=70$ than at $Re=50$ and $Re=60$. Moreover, the error at $Re=50$ exceeds that at $Re=60$, since only a single measurement location is used in that case, whereas for higher Reynolds numbers two measurement points of the $v$-velocity component are employed, leading to improved accuracy according to the chosen criterion.

The control is activated at $t=500$ and applied until $t=1000$. The MPC settings are identical to those used in the full-measurement case.

This time interval is sufficient to reduce $|\tilde{A}_p|$ to very small values, of order $10^{-7}$ for $Re=50$ and $Re=60$, with the corresponding amplitudes of the reduced input decreasing to values of order $10^{-8}$, similar as in the full measurement case.

For $Re=70$, control is more challenging due to the larger modelling errors. The amplitude of the reduced state initially increases, but is eventually reduced to approximately $10^{-4}$, with the corresponding input converging to values of order $10^{-5}$. 

The fluctuation energy evolution given in Figure~\ref{fig:energy} (dashed lines) confirms that the flow is effectively stabilized and brought to the steady base flow for $Re=50$ and $Re=60$. In both cases, the energy decreases steadily, albeit more slowly than in the full-measurement case, reaching values of order $10^{-5}$ at $t=1000$. A high level of suppression is also achieved for $Re=70$; although the energy initially increases, it is subsequently reduced to approximately $0.8$ by $t=1000$ and continues to decay under prolonged control, decreasing by about a factor of two over an additional $100$ time units (not shown here).
We do not show the vorticity fields again, as the differences between the full-measurement and point-measurement cases are not visually discernible.

\begin{figure}
\centering
\includegraphics[width=0.58\textwidth]{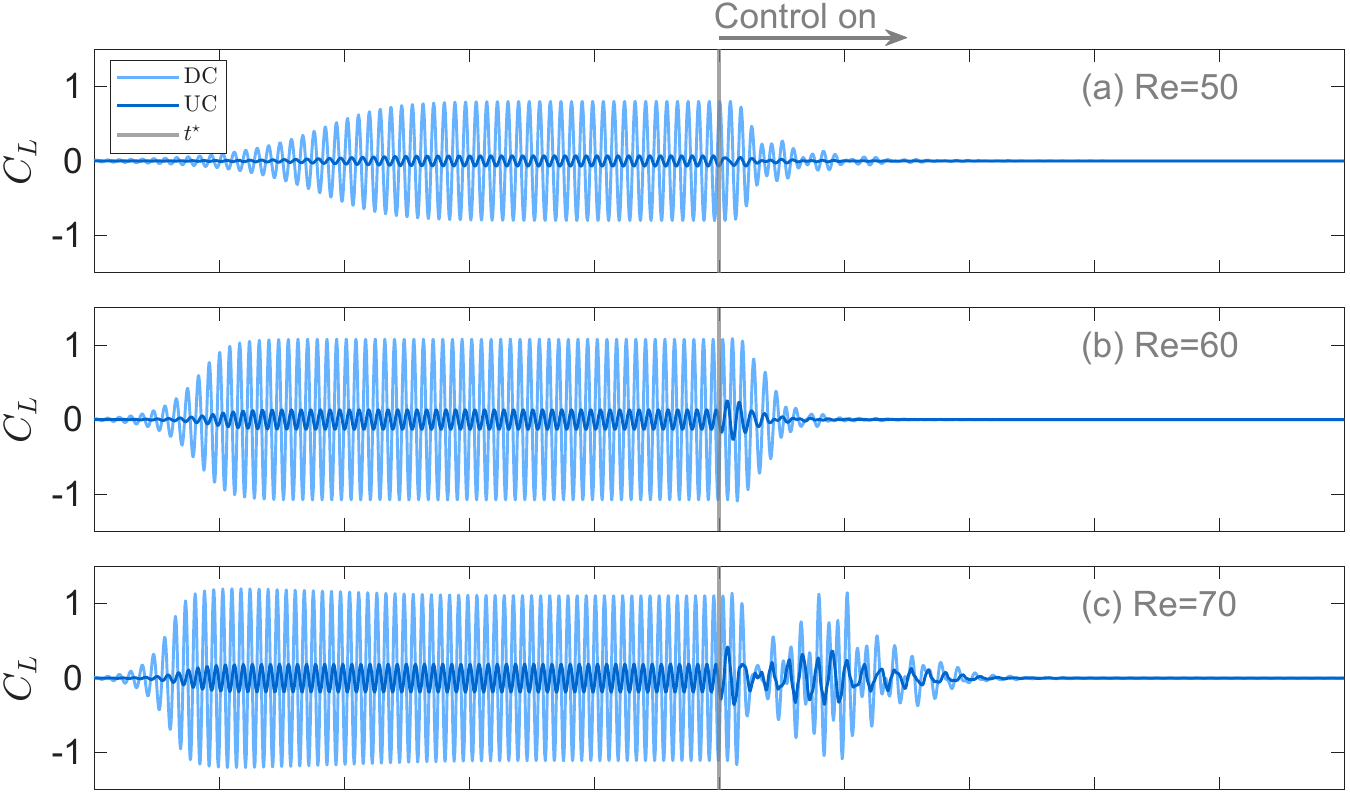}
\caption{Time evolution and suppression of lift coefficient $C_L$ acting on the upstream (UC) and downstream (DC) cylinders. The MPC feedback for the MPC is based on pointwise velocity measurements. (a) $Re=50$ (b) $Re=60$ (c) $Re=70$.}
\label{fig:CL_meas}
\end{figure}

\begin{figure}
\centering
\includegraphics[width=0.6\textwidth]{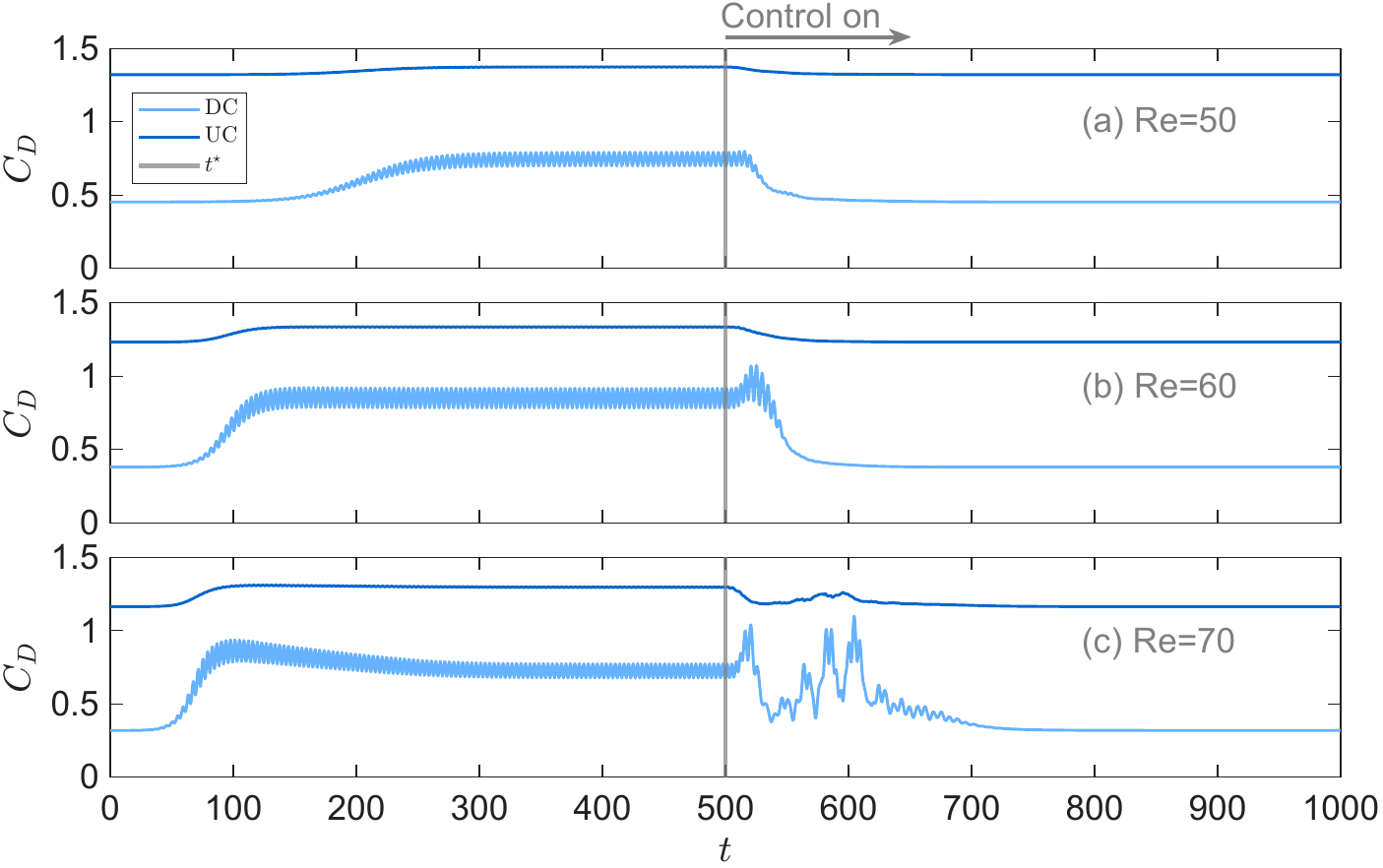}
\caption{Time evolution and suppression of drag coefficient $C_D$ acting on the upstream (UC) and downstream (DC) cylinders. The MPC feedback for the MPC is based on pointwise velocity measurements.(a) $Re=50$ (b) $Re=60$ (c) $Re=70$. }
\label{fig:CD_meas}
\end{figure}

Finally, we plot the time evolution of $C_L$ and $C_D$ in Figures~\ref{fig:CL_meas} and~\ref{fig:CD_meas}. 
Consistent with the vorticity observations, for $Re=50$, $60$, and $70$, the lift coefficient is reduced to near zero for both cylinders, reaching values of less than $10^{-6}$ for $Re=50$ and $Re=60$, and less than $10^{-3}$ for $Re=70$, while the drag coefficient approaches its steady value.

\section{Conclusion}

In this work, we have presented a model-based closed-loop control strategy for the flow over two cylinders in tandem configuration. Building on a weakly nonlinear analysis of the incompressible Navier--Stokes equations, we derived a low-dimensional, forced Stuart--Landau model that captures the essential nonlinear dynamics near the onset of instability, i.e.\ in the vicinity of the critical Reynolds number associated with a supercritical Hopf bifurcation. The formulation was extended to allow for time-dependent forcing amplitudes, enabling the design of feedback control laws.

Based on this reduced-order model, we designed a model predictive controller (MPC) to determine the control input in the reduced-order dynamics. The resulting control law is then mapped to the full-order system through explicit state and input mappings, yielding an output-feedback formulation. In this framework, the reduced state is estimated from available velocity measurements, and actuation is introduced through localized volumetric forcing to ensure physical realizability. Both full-field and pointwise measurement configurations are considered.

The proposed approach was demonstrated for tandem cylinders at spacing ratio $\gamma=8$, corresponding to the co-shedding regime, and Reynolds numbers $Re=50,60,70,$ and $80$. As expected from the weakly nonlinear expansion, the model accuracy decreases with increasing Reynolds number.

Using full-domain velocity measurements, the closed-loop controller successfully suppresses vortex shedding in both the gap region and the wake of the downstream cylinder for $Re=50, 60,$ and $70$, driving the flow close to the steady base state. For $Re=80$, the controller significantly mitigates the unsteady dynamics, leading to a reduced-amplitude oscillatory state. The control input required to achieve this suppression also approaches zero as the unsteadiness is suppressed, thereby achieving efficient control over long times.

We further demonstrated that effective control can be achieved using only a small number of pointwise velocity measurements. In particular, suppression is obtained for $Re=50$ using measurements at a single spatial location, and for $Re=60$ and $Re=70$ using two-point measurements, although the performance degrades compared to the full-measurement case as the Reynolds number increases.

Overall, these results demonstrate that the proposed model-based closed-loop control framework can effectively suppress or significantly reduce vortex shedding in tandem-cylinder flows using low-dimensional models, limited sensing, and localized actuation, while acknowledging the limitations associated with increasing Reynolds number and reduced model accuracy.

\section*{Acknowledgements}

A-J. Buchner was supported by the Netherlands Organisation for Scientific Research (NWO), under VENI project number 18176.






\appendix
\section{Weakly Nonlinear Analysis}\label{sec:WNAappendix}
Here we provide a detailed derivation of the Stuart--Landau equation~\eqref{eq:SL_resonance} for resonant forcing frequencies close to $\omega_0$. We also briefly summarize the corresponding results for other forcing frequencies, which can be obtained analogously.

\subsection{Forcing near the resonant frequency $ \omega_0$}\label{sec:resonantappendix}%

Substituting the weakly nonlinear expansion~\eqref{eq:expansion} and the forcing~\eqref{eq:forcing} into the governing equations~\eqref{eq:NS_compact}, together with the scaling~\eqref{eq:scaling_resonant}, and collecting terms at successive orders in $\sqrt{\epsilon}$ yields a hierarchy of linear inhomogeneous problems for $\mathbf{U}_1,\mathbf{U}_2,\mathbf{U}_3,\ldots$.
We solve these equations at each order $\sqrt{\epsilon}^{\,i}$ with $i=0,1,2,3$, and derive the Stuart--Landau equation~\eqref{eq:SL_resonance}.

\subsubsection{Order $\sqrt{\epsilon}^0$}
At zeroth order $O(\sqrt{\epsilon}^{\,0})$, we obtain the steady incompressible Navier--Stokes equations 
\begin{equation}\label{eq:base}
      \mathscr{N}(\mathbf{U}_0, Re_c) = \left(
\begin{array}{cc}
  -\nabla \mathbf{u}_0\cdot \mathbf{u}_0-\nabla p_0+ \frac{1}{Re_c}\Delta \mathbf{u}_0\\
  \displaystyle
    \nabla \cdot \mathbf{u}_0 \\
\end{array}  \right)=\mathbf{0}
\end{equation}
at the critical Reynolds number $Re_c$. By construction, this equation is satisfied by the base flow $\mathbf{U}_0$, which is an equilibrium of the unforced system at $Re_c$.

\subsubsection{Order $\sqrt{\epsilon}$}
At order $\sqrt{\epsilon}$, we obtain the the homogeneous linear problem
\begin{equation}\label{eq:NSlin_forced}
\Big(\frac{\partial}{\partial t} \mathscr{E} - \mathscr{A}_0 \Big)\mathbf{U}_1=\mathbf{0}, 
\end{equation}
where
\begin{equation*}
   \mathscr{A}_0=\left(
\begin{array}{cc}
   -\nabla () \cdot \mathbf{u}_0-\nabla \mathbf{u}_0\cdot ()+ \frac{1}{Re_c}\Delta &   \nabla  \\
  \displaystyle
   \nabla^T &  0 \\
\end{array}  \right)
\end{equation*}
denotes the linearized Navier--Stokes operator about the base flow $\mathbf{U}_0$. To express the general solution of~\eqref{eq:NSlin_forced} in terms of the eigenmodes of the operator pencil
\begin{equation*}
\mathscr{K}_{\lambda_0}=\lambda\mathscr{E}-\mathscr{A}_0,
\qquad \lambda_0=\sigma_0+\mathrm{i}\omega_0,
\end{equation*}
we consider the generalized eigenvalue problem
\begin{equation}\label{eq:eigenvalue}
\mathscr{K}_{\lambda_0}\hat{\mathbf{U}}=0.
\end{equation}

Since we assume that the first Hopf bifurcation occurs at $Re_c$, the spectrum
of $\mathscr{K}_{\lambda_0}$ contains a single marginally stable complex-conjugate eigenvalue pair $\pm \imath \omega_0$, with the associated eigenvector $\mathbf{U}_1^A$ (and its complex conjugate) satisfying
 \begin{equation*}\label{eq:order1}
     \mathscr{K}_{\imath \omega_0}\mathbf{U}_1^{A}=\mathbf{0},
 \end{equation*}
 while all remaining eigenvalues are stable. We therefore retain only the contribution along this eigenspace, which governs the long-term dynamics. The resulting leading-order solution can be written as
\begin{equation*}
    \mathbf{U}_{1}(t, \tau) = A(\tau) e^{\imath \omega_0 t}\mathbf{U}_1^A + c.c.,
  \end{equation*}
 where the complex amplitude $A(\tau)$ evolves on the slow time scale. The mode $\mathbf{U}_1^{A}$ is referred to as the \emph{critical global mode}~\cite{sipp2007global}, and the term $A(\tau)e^{\imath \omega_0 t}\mathbf{U}_1^A$ represents the first harmonic, oscillating at the fundamental frequency $\omega_0$ on the fast time scale.

\subsubsection{Order $\sqrt{\epsilon}^{\,2}$}
At order $\sqrt{\epsilon}^{\,2}$, we obtain the linear inhomogeneous problem
\begin{eqnarray*}
       \Big(\frac{\partial}{\partial t} \mathscr{E} - \mathscr{A}\Big)\mathbf{U}_2&=&\mathbf{F}_2^1+\lvert A (\tau)\rvert^2 \mathbf{F}^{\lvert A \rvert^2} +\big(A(\tau)^2 e^{\imath 2\omega_0t}\mathbf{F}^{A^2}+c.c.\big),
\end{eqnarray*}
where the right-hand side terms are
\begin{align*}
    \mathbf{F}^{\lvert A \rvert^2}&:=\mathscr{P}\big(
     -\nabla \mathbf{u}_1^A \cdot \overline{\mathbf{u}_1^A}-\nabla \overline{\mathbf{u}_1^A} \cdot \mathbf{u}_1^A \big),~
     \mathbf{F}^{A^2}:=\mathscr{P}\big(
     -\nabla \mathbf{u}_1^A \cdot \mathbf{u}_1^A  \big), \\
     \mathbf{F}_2^1&:=\mathscr{P}\big(
        - \Delta \mathbf{u}_0 \big).
\end{align*}
Accordingly, we write $\mathbf{U}_2(t,\tau)$ as a superposition of the responses to each right-hand-side contribution,
\begin{align}
       \mathbf{U}_2(t, \tau) & = \mathbf{U}_2^1+\lvert A (\tau) \rvert^2 \mathbf{U}_2^{\lvert A \rvert^2} +A(\tau)^2 e^{\imath 2\omega_0t}\mathbf{U}_2^{A^2}+c.c.,\label{eq:u2_forced}
\end{align}
where the fields $\mathbf{U}_2^1$, $\mathbf{U}_2^{|A|^2}$, and $\mathbf{U}_2^{A^2}$ are obtained from
\begin{align} \label{eq:order2}
    \mathscr{K}_0 \mathbf{U}_2^1&= \mathbf{F}_2^1,\quad
    \mathscr{K}_0 \mathbf{U}_2^{\lvert A \rvert^2}=\mathbf{F}^{\lvert A \rvert^2}, \quad
    \mathscr{K}_{\imath 2\omega_0} \mathbf{U}_2^{ A^2}=\mathbf{F}^{A^2}. 
\end{align}
By the assumptions of the Hopf bifurcation theorem (see Theorem 8.25 in~\citet{chicone2006ordinary}), $\lambda=0$ and $\lambda=\imath 2\omega_0$ do not belong to the spectrum of the operator pencil $\mathscr{K}_{\lambda_0}$, and therefore the problems in~\eqref{eq:order2} admit unique solutions.

The component $\mathbf{U}_2^1$ represents the $O(\epsilon)$ correction to $\mathbf{U}_0$ that accounts for the shift of the unstable equilibrium branch for $\epsilon>0$. The term $A(\tau)^2 e^{\imath 2\omega_0 t}\mathbf{U}_2^{A^2}$ is the second harmonic, oscillating at twice the fundamental frequency, and stems from the nonlinear interaction between the first harmonic with itself.  The mean-flow component $|A(\tau)|^2\,\mathbf{U}_2^{|A|^2}$ results from the interaction of the first harmonic with its complex conjugate; it evolves on the slow time scale and approximates the difference between the mean flow and the steady base flow~\cite{sipp2007global}.

\subsubsection{Order $\sqrt{\epsilon}^{\,3}$}
At order $\sqrt{\epsilon}^{\,3}$, we obtain the inhomogeneous linear problem 
\begin{align}\label{eq:thirdorder}
     \Big(\frac{\partial}{\partial t} \mathscr{E}-\mathscr{A} \Big)\mathbf{U}_3 = &-\mathscr{E} \frac{d A}{d \tau} e^{\imath \omega_0 t} \mathbf{U}_1^A+A (\tau) e^{\imath \omega_0 t} \mathbf{F}^A\nonumber \\
     &+A (\tau)\lvert A(\tau) \rvert^2 e^{\imath \omega_0 t} \mathbf{F}^{A\lvert A \rvert^2}  +E(\tau) e^{\imath \omega_f t} \mathcal{P}\mathbf{f}_E \nonumber \\
     & +c.c.+ ...
\end{align}
where the right-hand side terms $\mathbf{F}^A$ and $\mathbf{F}^{A \lvert A \rvert^2}$ are defined by
\begin{align*}
           \mathbf{F}^A & :=\mathscr{P}\big(
          -\nabla \mathbf{u}_1^A \cdot \mathbf{u}_2^1- \nabla \mathbf{u}_2^1 \cdot \mathbf{u}_1^A- \Delta \mathbf{u}_1^A \big), \nonumber \\
         \mathbf{F}^{A \lvert A \rvert^2}& :=\mathscr{P} \big(-\nabla \mathbf{u}_1^A \cdot \mathbf{u}_2^{\lvert A \rvert^2}-\nabla \mathbf{u}_2^{\lvert A \rvert^2}\cdot \mathbf{u}_1^A-\nabla \bar{\mathbf{u}}_1^A \cdot \mathbf{u}_2^{A^2}-\nabla \mathbf{u}_2^{A^2} \cdot \bar{\mathbf{u}}_1^A \big). 
\end{align*}
The term $\mathbf{F}^A$ stems from the viscous diffusion of the first harmonic $Ae^{\imath \omega_0 t}\mathbf{U}_1^{A}$ and its interaction with the $O(\epsilon)$ base-flow correction $\mathbf{U}_2^{1}$. The term $\mathbf{F}^{A\lvert A \rvert^2}$ results from interactions between the first harmonic
$A (\tau)e^{\imath \omega_0 t}\mathbf{U}_1^{A}$, the zeroth harmonic $\lvert A(\tau) \rvert^2 \mathbf{U}_2^{\lvert A \rvert^2}$, and the second harmonic $A(\tau)^2 e^{\imath 2\omega_0t}\mathbf{U}_2^{A^2}$.
Since the right-hand side of~\eqref{eq:thirdorder} contains contributions oscillating at (or close to) the marginal frequency $\omega_0$, the corresponding particular solution $\mathbf{U}_3(t,\tau)$ may exhibit secular growth in time unless appropriate compatibility conditions are enforced.




\subsubsection{Stuart-Landau Equation}

To ensure that the solution remains bounded in time, and hence that the asymptotic expansion remains valid, the resonant terms on the right-hand side of~\eqref{eq:thirdorder} must satisfy a compatibility condition. In particular, they must be orthogonal, at all times, to the adjoint eigenvector $\mathbf{U}_1^{A\star}$ associated with $\mathscr{K}_{\imath\omega_0}$. This condition imposes an evolution equation for the global-mode amplitude $A$, which takes the Stuart--Landau form

\begin{equation}\label{eq:SL_resonance_t_appendix}
    \phantom{.}\frac{d A}{d \tau}=a_0 A - a_1 A \lvert A \rvert^2 +a_2 e^{\imath \epsilon \Omega t}E, 
\end{equation}
%
%
where the coefficients $a_0$, $a_1$, and $a_2$ are obtained from the corresponding compatibility (orthogonality) conditions
\begin{subequations}
\label{eq:SLcoefficients}
\begin{eqnarray}
   a_0&:=&\frac{\langle \mathbf{U}_1^{A \star},\mathbf{F}^A  \rangle}{\langle \mathbf{U}_1^{A \star},  \mathscr{E} \mathbf{U}_1^A \rangle} \label{eq:a0} \\
     a_1&:=&-\frac{\langle \mathbf{U}_1^{A \star},\mathbf{F}^{A \lvert A \rvert^2 } \rangle}{\langle \mathbf{U}_1^{A \star},  \mathscr{E} \mathbf{U}_1^A \rangle} \label{eq:a1} \\
a_2&:=&\frac{\langle \mathbf{U}_1^{A \star},\mathscr{P}\mathbf{f}_E \rangle}{\langle \mathbf{U}_1^{A \star},  \mathscr{E} \mathbf{U}_1^A \rangle}.\label{eq:a2}
\end{eqnarray}
\end{subequations}
Here, the scalar product between two fields $\mathbf{U}_{\alpha}$ and $\mathbf{U}_{\beta}$ is defined as
\begin{equation}\label{eq:scalar_product}
    \langle \mathbf{U}_{\alpha}, \mathbf{U}_{\beta} \rangle := \iint_{\Omega} (\bar{u}_{\alpha}u_{\beta}+\bar{v}_{\alpha}v_{\beta}+\bar{p}_{\alpha}p_{\beta}) dx dy.
\end{equation}
Thus, the solution at order $\sqrt{\epsilon}^{\,2}$ takes the form 
\begin{align*}
    \mathbf{U}_3(t, \tau) & =A(\tau)e^{\imath \omega_0 t}\mathbf{U}_3^A+A(\tau)\lvert A(\tau) \rvert^2 e^{\imath \omega_0 t}\mathbf{U}_3^{A\lvert A \rvert^2} \\
    & \quad +E(\tau)e^{\imath \omega_f t}\mathbf{U}_3^E+c.c.+ ...
\end{align*}
where we retain only the contributions oscillating at (or close to) the resonant frequency.

\subsection{Other Resonant Forcing Frequencies}\label{sec:otherappendix}
We next discuss other resonant forcing cases, namely $\omega_f=0$, $\omega_f\approx 2\omega_0$, and $\omega_f\approx \omega_0/2$. Each case requires a different forcing scaling in order for the forcing to enter the Stuart--Landau equation at $O(\sqrt{\epsilon}^{\,3})$, while avoiding degenerate (non-invertible) operators at lower orders.

\subsubsection{Forcing at $\omega_f=0$}
For $\omega_f=0$, the forcing amplitude is scaled as
\begin{equation*}
    E'(\tau):= \epsilon E (\tau).
\end{equation*}
Solving the resulting equations at orders $\sqrt{\epsilon}^0$, $\sqrt{\epsilon}^1$, $\sqrt{\epsilon}^2$ and $\sqrt{\epsilon}^3$, yields
\begin{align}\label{eq:wna_zero}
    \mathbf{U} \approx\; & \mathbf{U}_0+ \sqrt{\epsilon} \big(A e^{\imath \omega_0 t}\mathbf{U}_1^A + c.c.\big)
\nonumber \\ 
 & +\epsilon \big( \mathbf{U}_2^1+\lvert A \rvert^2 \mathbf{U}_2^{\lvert A \rvert^2} + \big(A^2 e^{\imath 2\omega_0t}\mathbf{U}_2^{A^2}+E \mathbf{U}_2^E + c.c.\big) \big)\nonumber \\
     & +\sqrt{\epsilon}^{\,3} \big(A e^{\imath \omega_0 t} \mathbf{U}_3^A+A\lvert A \rvert^2 e^{\imath \omega_0 t} \mathbf{U}_3^{A\lvert A \rvert^2} \nonumber \\
     & \quad +AE e^{\imath \omega_0 t} \mathbf{U}_3^{AE} +
     A\bar{E} e^{\imath \omega_0 t} \mathbf{U}_3^{A\bar{E}} + c.c.\big) + ..., 
\end{align} 
where $\mathbf{U}_0$, $\mathbf{U}_1^A$, $\mathbf{U}_2^1$, $\mathbf{U}_2^{\lvert A \rvert^2}$, $\mathbf{U}_2^{A^2}$ are identical to those in Sec.~\ref{sec:resonantappendix}, while the
forcing response $\mathbf{U}_2^E$ is obtained from
\begin{align*}
\mathscr{K}_{0}\mathbf{U}_2^{E}&=\mathscr{P}  \mathbf{f}_E.
\end{align*} 
At order $\sqrt{\epsilon}^{\,3}$, the contributions $\mathbf{U}_3^A$, $\mathbf{U}_3^{A\lvert A \rvert^2}$, $\mathbf{U}_3^{A E}$ and $\mathbf{U}_3^{A \bar{E}}$ must again satisfy compatibility conditions. These yield the Stuart--Landau equation
\begin{equation}\label{eq:SL_omega0}
    \frac{d A}{d t}=\epsilon(a_0-a_2E-a_3\bar{E})A- \epsilon a_1 A \lvert A \rvert^2.
    \end{equation}
    Since the coefficients $a_0$ and $a_1$ depend only on the unforced modes, they are independent of the forcing and are given, as in Secion~\ref{sec:resonantappendix}, by~\eqref{eq:a0} and~\eqref{eq:a1}. The forcing-related coefficient is
 
\begin{align*}
 a_2&:=\frac{\langle \mathbf{U}_1^{A \star}, \mathscr{P}\big( \nabla \mathbf{u}_1^A \cdot \mathbf{u}_1^E+\nabla \mathbf{u}_1^E\cdot \mathbf{u}_1^A 
     \big) \rangle}{\langle \mathbf{U}_1^{A \star},  \mathscr{E} \mathbf{U}_1^A \rangle} \\
      a_3&:=\frac{\langle \mathbf{U}_1^{A \star},\mathscr{P} \big( \nabla \mathbf{u}_1^A \cdot \overline{\mathbf{u}_1^E}+\nabla \overline{\mathbf{u}_1^E}\cdot \mathbf{u}_1^A 
     \big) \rangle}{\langle \mathbf{U}_1^{A \star},  \mathscr{E} \mathbf{U}_1^A \rangle}.
\end{align*}
Equation~\eqref{eq:SL_omega0} shows that, for $\omega_f=0$, the forcing enters the reduced dynamics \emph{bilinearly} through the term $(a_2E+a_3\bar{E})A$, in contrast to the purely additive forcing term in~\eqref{eq:SL_resonance_t_appendix}.

\subsubsection{Forcing near $2\omega_0$}
For a forcing frequency near $2\omega_0$, namely
$\omega_f = 2\omega_0+\epsilon \Omega$ for some frequency $\Omega$, we scale the forcing amplitude as
\begin{equation*}
    E'(\tau):= \epsilon\,E(\tau).
\end{equation*}

Solving the resulting equations yields an expansion of the form
\begin{align}\label{eq:wna_double}
    \mathbf{U} \approx ~\mathbf{U}_0&+\sqrt{\epsilon} \big(A e^{\imath \omega_0 t}\mathbf{U}_1^A + c.c.\big)
\nonumber \\\nonumber
 & +\epsilon \big(\mathbf{U}_2^1+\lvert A \rvert^2 \mathbf{U}_2^{\lvert A \rvert^2} \\ \nonumber
 & + \big(A^2 e^{\imath 2\omega_0t}\mathbf{U}_2^{A^2} +E e^{\imath (2\omega_0+\epsilon \Omega)t}\mathbf{U}_2^E + c.c.\big )\big) \nonumber \\
     & +\sqrt{\epsilon}^{\,3} \big( A e^{\imath \omega_0 t} \mathbf{U}_3^A+A\lvert A \rvert^2 e^{\imath \omega_0 t} \mathbf{U}_3^{A\lvert A \rvert^2}\\ \nonumber
    & +
     \bar{A}E e^{\imath (\omega_0+\epsilon \Omega) t} \mathbf{U}_3^{\bar{A}E} + c.c. \big) + .... 
\end{align} 

Here, the forcing response is obtained from
\begin{align*}
\mathscr{K}_{\imath (2\omega_0+\epsilon \Omega)}\mathbf{U}_2^{E}&=\mathscr{P}  \mathbf{f}_E.
\end{align*} 
The corresponding Stuart--Landau equation is 
\begin{equation*}
  \frac{d A}{d t}=\epsilon a_0 A- \epsilon a_1  A \lvert A \rvert^2 -\epsilon a_2 e^{\imath \epsilon \Omega t}\bar{A}E
\end{equation*}
with the coefficient 
\begin{align*}
 a_2:=\frac{\langle \mathbf{U}_1^{A \star}, \mathscr{P}\big( \nabla \overline{\mathbf{u}_1^A} \cdot \mathbf{u}_1^E+ \nabla \mathbf{u}_1^E \cdot \overline{\mathbf{u}_1^A} 
     \big) \rangle}{\langle \mathbf{U}_1^{A \star},  \mathscr{E} \mathbf{U}_1^A \rangle}
\end{align*}
is determined by the interaction between the forcing response and the complex conjugate of the global mode.

\subsubsection{Forcing near $\omega_0/2$}
The last resonant forcing case concerns frequencies near $\omega_0/2$. We assume
 $\omega_f=\omega_0/2+\epsilon \Omega$. The appropriate scaling of $E'(\tau)$ is
\begin{equation*}
    E'(\tau):= \epsilon^{3/4}  E(\tau)
\end{equation*}
and the weakly nonlinear expansion takes the form
\begin{align}\label{eq:wna_half}
    \mathbf{U} \approx ~\mathbf{U}_0&+\sqrt{\epsilon}\big(A e^{\imath \omega_0 t}\mathbf{U}_1^A + c.c.\big)
\nonumber \\ 
& +\sqrt{\epsilon}^{\,3/2} \big(E e^{\imath (\omega_0/2+\epsilon \Omega) t} \mathbf{U}_2^E + c.c. \big) \nonumber \\
 & +\epsilon \big(\mathbf{U}_2^1+\lvert A \rvert^2 \mathbf{U}_2^{\lvert A \rvert^2} +A^2 e^{\imath 2\omega_0t}\mathbf{U}_2^{A^2} \big)\nonumber \\
     & +\sqrt{\epsilon}^{\,3} \big(A e^{\imath \omega_0 t} \mathbf{U}_3^A+A\lvert A\rvert^2 e^{\imath \omega_0 t} \mathbf{U}_3^{A\lvert A \rvert^2} \nonumber \\
     & \quad +
     E^2 e^{\imath (\omega_0+2 \epsilon \Omega) t} \mathbf{U}_3^{E^2} + c.c.\big) + ..., 
\end{align} 
where the forcing response $\mathbf{U}_2^{E}$ is obtained from
\begin{align*}
\mathscr{K}_{\imath (\omega_0/2+ \epsilon \Omega)}\mathbf{U}_2^{E}&=\mathscr{P}  \mathbf{f}_E. 
\end{align*} 
The corresponding Stuart--Landau equation is

\begin{equation}\label{eq:SL_omegahalf}
  \frac{d A}{d t}=\epsilon a_0 A- \epsilon  a_1 A \lvert A \rvert^2 - \epsilon a_2 e^{\imath 2 \epsilon \Omega}E^2,
\end{equation}
where
\begin{align*}
 a_2:=\frac{\langle \mathbf{U}_1^{A \star}, \mathscr{P}\big(\nabla \mathbf{u}_1^E \cdot \mathbf{u}_1^E
     \big) \rangle}{\langle \mathbf{U}_1^{A \star},  \mathscr{E} \mathbf{U}_1^A \rangle} 
\end{align*}
stems from the nonlinear self-interaction of the forcing response. In this case, the forcing enters~\eqref{eq:SL_omegahalf} as a nonlinear additive term, in analogy to~\eqref{eq:SL_resonance_t_appendix} for forcing near the natural frequency.

\subsection{Non-Resonant Forcing Frequencies}\label{sec:nonresonant}
For non-resonant forcing, i.e.\ for $\omega_f\neq 0$ and $\omega_f \not\approx \omega_0,\ \omega_0/2,\ 2\omega_0$,
the forcing amplitude scales as
\begin{equation}\label{eq:scaling_nonresonant}
    E'(\tau):=\sqrt{\epsilon}E(\tau).
\end{equation}
The equations at orders $\sqrt{\epsilon}^{\,0}$, $\sqrt{\epsilon}^{\,1}$, $\sqrt{\epsilon}^{\,2}$, and $\sqrt{\epsilon}^{\,3}$ are solved successively as in the resonant cases, yielding the expansion
\begin{align}\label{eq:wna_nonres}
    \mathbf{U} \approx ~\mathbf{U}_0&+ \sqrt{\epsilon} \big(A e^{\imath \omega_0 t}\mathbf{U}_1^A+E e^{\imath \omega_f t}\mathbf{U}_1^E + c.c.\big) 
\nonumber \\ 
     & +\epsilon \big(\mathbf{U}_2^1+\lvert A \rvert^2 \mathbf{U}_2^{\lvert A \rvert^2} + \big(A^2 e^{\imath 2\omega_0t}\mathbf{U}_2^{A^2}+ c.c.\big) \nonumber  \\
&+\lvert E \rvert^2 \mathbf{U}_2^{\lvert E \rvert^2} +\big(E^2 e^{\imath 2\omega_f t}\mathbf{U}_2^{E^2}+c.c.\big) \nonumber  \\
       &+\big(AEe^{\imath (\omega_0+\omega_f)t}\mathbf{U}_2^{AE}+ A\bar{E}e^{\imath (\omega_0-\omega_f)t}\mathbf{U}_2^{A\bar{E}}+c.c.\big) \big) \nonumber \\
     & +\sqrt{\epsilon}^{\,3} \big( A e^{\imath \omega_0 t} \mathbf{U}_3^A+A\lvert A \rvert^2 e^{\imath \omega_0 t} \mathbf{U}_3^{A\lvert A \rvert^2} \nonumber \\
     &\quad +A\lvert E\rvert^2 e^{\imath \omega_0 t} \mathbf{U}_3^{A \lvert E\rvert^2}+ c.c.\big)+ ... 
\end{align} 
The fields $\mathbf{U}_1^E$, $\mathbf{U}_2^{|E|^2}$, $\mathbf{U}_2^{E^2}$, $\mathbf{U}_2^{AE}$, and $\mathbf{U}_2^{A\bar{E}}$ are obtained by solving

\begin{align*}
\mathscr{K}_{\imath \omega_f}\mathbf{U}_1^{E}&=\mathscr{P}  \mathbf{f}_E, \\
 \mathscr{K}_{0}\mathbf{U}_2^{\lvert E \rvert^2} &=   \mathscr{P}\big(
         -\nabla \mathbf{u}_1^E \cdot \overline{\mathbf{u}_1^E}-\nabla \overline{\mathbf{u}_1^E} \cdot \mathbf{u}_1^E \big),\\
      \mathscr{K}_{2\imath \omega_f}\mathbf{U}_2^{E^2} &=   \mathscr{P}\big(
         -\nabla \mathbf{u}_1^E \cdot \mathbf{u}_1^E \big),   \\
    \mathscr{K}_{\imath (\omega_0+\omega_f)}\mathbf{U}_2^{AE} &= \mathscr{P}\big( -\nabla \mathbf{u}_1^A \cdot \mathbf{u}_1^E-\nabla \mathbf{u}_1^E \cdot \mathbf{u}_1^A 
     \big), \\
    \mathscr{K}_{\imath (\omega_0-\omega_f)}\mathbf{U}_2^{A\bar{E}} &= \mathscr{P}\big( -\nabla \mathbf{u}_1^A \cdot \overline{\mathbf{u}_1^E}-\nabla \overline{\mathbf{u}_1^E}\cdot \mathbf{u}_1^A 
     \big).
\end{align*} 

As in the previous cases, at order $\sqrt{\epsilon}^{\,3}$, the terms $\mathbf{U}_3^A$, $\mathbf{U}_3^{A\lvert A \rvert^2}$ and $\mathbf{U}_3^{A \lvert E\rvert^2}$ need to satisfy compatibility conditions which yield the Stuart-Landau equation 
\begin{equation*}
    \frac{d A}{d t}=\epsilon (a_0- a_2 \lvert E \rvert^2) A - \epsilon  a_1 A \lvert A\rvert^2 
\end{equation*}
with 
\begin{equation*}
   a_2:=\frac{\big\langle \mathbf{U}_1^{A \star},\mathscr{P}\big(\nabla \mathbf{u}_1^A \cdot \mathbf{u}_2^{ \lvert E\rvert^2}+\nabla \mathbf{u}_2^{ \lvert E\rvert^2} \cdot \mathbf{u}_1^A  +\nabla \mathbf{u}_1^E \cdot \mathbf{u}_2^{A \bar{E}}+\nabla \mathbf{u}_2^{A \bar{E}} \cdot \mathbf{u}_1^E \big) \big\rangle}{\langle \mathbf{U}_1^{A \star},  \mathscr{E} \mathbf{U}_1^A  \rangle}.      
\end{equation*}
As in the cases of zero-frequency forcing and forcing near $2\omega_0$, the input acts multiplicatively on the linear part of the reduced dynamics.

\subsection{Stabilization Capabilities across Different Frequencies} \label{sec:stabilization}

As suggested by the weakly nonlinear approximations of $\mathbf{U}$ given in~\eqref{eq:solution_wna1},~\eqref{eq:wna_zero},~\eqref{eq:wna_double},~\eqref{eq:wna_half}, and~\eqref{eq:wna_nonres}, the equilibrium
$\mathbf{U}_b\approx \mathbf{U}_0+\epsilon\,\mathbf{U}_2^1$ corresponds to $A=0$ and $E=0$.

In the case of zero-frequency forcing, forcing near $2 \omega_0$, and non-resonant frequency, the forcing amplitude $E$ cannot converge to zero
while maintaining stability of the Stuart-Landau model as it acts multiplicatively on the linear part of its dynamics.
This implies the following. First, to stabilize the Stuart-Landau model, we need a persisting control input, which can be energy inefficient. Second, although we might stabilize the Stuart-Landau model, and therefore the global mode and all the other unsteady terms in~\eqref{eq:wna_zero}, ~\eqref{eq:wna_double},~\eqref{eq:wna_nonres} that arise from its interaction with itself or the forcing response, terms that depend exclusively on the forcing persist. For unsteady forcing with frequency near $2 \omega_0$ or a non-resonant frequency, this implies that the flow will remain unsteady.

However, for the cases of $\omega_f \approx \omega_0$ and $\omega_f \approx \omega_0/2$, the
amplitude $E$ enters the Stuart-Landau equation as an additive term which allows bringing $E$ to zero while stabilizing the model. Once both $A$ and $E$ converge to zero, the flow is at its equilibrium $\mathbf{U}_b\approx \mathbf{U}_0+\epsilon\mathbf{U}_2^1$ according to~\eqref{eq:solution_wna1} and~\eqref{eq:wna_half}. Therefore, the unsteadiness of the flow can be fully suppressed.

We choose forcing near the resonant frequency $\omega_f \approx \omega_0$ since this case facilitates characterizing the structure $f_E$ which optimizes the control efficiency, 
as shown in Section~\ref{sec:optimalforcing}. Finding such a forcing structure for the forcing frequency $\omega_f \approx \omega_0/2$ is a nontrivial task and we leave it for future research.

\section{Mesh Convergence}\label{Meshconvergence}

We assess the convergence of the coefficients obtained from the weakly nonlinear analysis with respect to the computational mesh. For a consistent comparison, the critical global mode is normalized such that $v_1^A(0.65,0) = -0.0979 + 0.2661\imath$ for all meshes. The meshes are unstructured and generated using the adaptive meshing procedure \texttt{AdaptMesh} in \texttt{FreeFEM++}, based on a Delaunay--Voronoi algorithm. In all cases, the mesh is adapted to the base flow as well as to the structure of the direct and adjoint global modes.

\begin{table}
  \begin{center}
\def~{\hphantom{0}}
  \begin{tabular}{lcccccccc}
Mesh & $Nt$  & $h_{min}$ & $\lambda_0$   &  $a_0$  &  $a_1$\\ [5pt] \hline
 M1 &  $490896$& $0.0047$& $-5.9\times 10^{-5}+0.6747\imath$& $ 7.6888+2.8863\imath$ & $7.6883-30.1099\imath$ \\
 M2 & $359028$ & $0.0528$& $-0.0004+0.6756\imath$ &$7.6571+2.9857\imath$ & $7.1153-31.2244\imath$ \\
 M3 &$253344$ &$0.0577$&$-0.0005+0.6756\imath$ & $7.5126+3.1444\imath$ & $8.3083-36.4366\imath$\\
 M1bd &  $537456$& $0.0038$& $5.9\times 10^{-5}+0.6747\imath$& $ 7.6896+2.8860\imath$ & $7.6457-29.9435\imath$ \\
  \end{tabular}
   \caption{Values of the coefficients of the Stuart-Landau model obtained via weakly nonlinear analysis (WNA). The coefficient $a_2$ is
calculated for the localized forcing structure~\eqref{eq:localforcing}.}
   \label{tab:SLcoeff_mesh}
  \end{center}
\end{table}

We consider three meshes, M1, M2, and M3, defined on the same computational domain (see Section~\ref{sec:computationaldomain}), which differ only in the minimal element size and, consequently, in the number of elements. The results reported in Table~\ref{tab:SLcoeff_mesh} show that mesh M2 provides sufficiently accurate estimates of the weakly nonlinear coefficients while being computationally more efficient than the finer mesh M1. In contrast, further coarsening to mesh M3 leads to more noticeable deviations, particularly in the nonlinear coefficient $a_1$.

We also examine the influence of the computational domain extent. The upper and lower boundaries are fixed at $y=\pm 30$, following~\citet{sipp2007global}, where it was shown for the single-cylinder case that smaller vertical extents introduce slight blockage effects. 
To assess the sensitivity to the downstream boundary, mesh M1bd uses the same resolution as M1 but extends the outlet to $x_{+\infty}=250$. As shown in Table~\ref{tab:SLcoeff_mesh}, this modification has a negligible effect on the computed coefficients, indicating that the original domain size is sufficient to capture the relevant flow dynamics and that the parameter estimation is not sensitive to the outlet position.


\section{Code Validation}\label{sec:codevalidation}

To validate the DNS carried out in this paper, we compare the hydrodynamic loads experienced by the cylinders, as well as the Strouhal number $St$ of vortex shedding, against available results in the literature. Since reference data are available at $Re=100$, we perform this validation at this Reynolds number. 

Specifically, we compare the root mean square (RMS) lift and drag coefficients, $C_L'$ and $C_D'$, together with the mean drag coefficient $\bar{C}_D$, for the fully developed unforced flow at the limit cycle. The results are summarized in Table~\ref{tab:table3} and compared with the numerical data of~\citet{sharman2005numerical} for $\gamma=8$ at $Re=100$.

\begin{table}
\centering
\caption{\label{tab:table3}  Comparison of DNS results with available literature~\cite{sharman2005numerical} for $\gamma=8$ at $Re=100$. $C_l'$ and $C_d'$ here refer to RMS value of lift and drag coefficient at the limit cycle, $C_d$ is the mean drag coefficient at the limit cycle and $St$ is Strouhal number.}

\begin{tabular}{ccc}
 Variable &Present& Literature\\ \hline 
$C_l'$ (UC)& 0.2332 & 0.2377  \\
$C_l'$ (DC)& 0.7550 &   0.7760\\
$C_d$ (UC) &1.2552 & 1.2966  \\
$C_d$ (DC)& 0.5876&   0.6031 \\
$C_d'$ (UC) & 0.0067& 0.0066 \\
$C_d'$ (DC)& 0.0480&   0.0501 \\
$St$& 0.15625 &  0.1529
\\
\end{tabular}

\end{table}
We observe good agreement between the present results and those reported in the literature.

In addition, we compare the critical Reynolds number obtained from our analysis, $Re_c=44.1$, with the value reported in~\citet{wang2022first}, further supporting the accuracy of the numerical framework.

\bibliographystyle{plainnat}
\bibliography{ref}

@article{sumner2010two,
  title={Two circular cylinders in cross-flow: A review},
  author={Sumner, D},
  journal={Journal of fluids and structures},
  volume={26},
  number={6},
  pages={849--899},
  year={2010},
  publisher={Elsevier}
}

@article{zdravkovich1987effects,
  title={The effects of interference between circular cylinders in cross flow},
  author={Zdravkovich, M.M.},
  journal={Journal of Fluids and Structures},
  volume={1},
  number={2},
  pages={239--261},
  year={1987},
  publisher={Elsevier}
}

@article{xu2004strouhal,
  title={Strouhal numbers in the wake of two inline cylinders},
  author={Xu, G and Zhou, Y},
  journal={Experiments in Fluids},
  volume={37},
  pages={248--256},
  year={2004},
  publisher={Springer}
}

@article{liu2015tandem,
  title={Tandem cylinder aerodynamic sound control using porous coating},
  author={Liu, H.and Azarpeyvand, M. and Wei, J. and Qu, Z.},
  journal={Journal of Sound and Vibration},
  volume={334},
  pages={190--201},
  year={2015},
  publisher={Elsevier}
}

@article{noack2003hierarchy,
  title={A hierarchy of low-dimensional models for the transient and post-transient cylinder wake},
  author={Noack, B.R. and Afanasiev, K. and Morzy{\'n}ski, M. and Tadmor, G. and Thiele, F.},
  journal={Journal of Fluid Mechanics},
  volume={497},
  pages={335--363},
  year={2003},
  publisher={Cambridge University Press}
}

@article{zhou2006flow,
  title={Flow structure, momentum and heat transport in a two-tandem-cylinder wake},
  author={Zhou, Y. and Yiu, M.W.},
  journal={Journal of Fluid Mechanics},
  volume={548},
  pages={17--48},
  year={2006},
  publisher={Cambridge University Press}
}

@article{kozlov2011plasma,
  title={Plasma flow control of cylinders in a tandem configuration},
  author={Kozlov, A.V. and Thomas, F.O.},
  journal={AIAA journal},
  volume={49},
  number={10},
  pages={2183--2193},
  year={2011}
}

@article{eltaweel2014numerical,
  title={Numerical investigation of tandem-cylinder noise reduction using plasma-based flow control},
  author={Eltaweel, A. and Wang, M. and Kim, D. and Thomas, F.O. and Kozlov, A.V.},
  journal={Journal of Fluid Mechanics},
  volume={756},
  pages={422--451},
  year={2014},
  publisher={Cambridge University Press}
}

@article{wolfe2003feedback,
  title={Feedback control of vortex shedding from two tandem cylinders},
  author={Wolfe, D. and Ziada, S.},
  journal={Journal of Fluids and Structures},
  volume={17},
  number={4},
  pages={579--592},
  year={2003},
  publisher={Elsevier}
}

@article{sipp2007global,
  title={Global stability of base and mean flows: a general approach and its applications to cylinder and open cavity flows},
  author={Sipp, D. and Lebedev, A.},
  journal={Journal of Fluid Mechanics},
  volume={593},
  pages={333--358},
  year={2007},
  publisher={Cambridge University Press}
}

@inbook{fauve2009,
    author = "S. Fauve",
    title = "Hydrodynamics and Nonlinear Instabilities",
    publisher = "Cambridge University Press",
    year = "2009",
    chapter = "Pattern Forming Instabilities"
}

@article{sipp2012open,
  title={Open-loop control of cavity oscillations with harmonic forcings},
  author={Sipp, D.},
  journal={Journal of Fluid Mechanics},
  volume={708},
  pages={439--468},
  year={2012},
  publisher={Cambridge University Press}
}

@book{vandyke75perturbation,
    author = {{van~Dyke}, M.},
    title = {Perturbation Methods in Fluid Mechanics},
    publisher = {Parabolic Press, Stanford,
CA},
    year =1975,
}

@article{jin2022resolvent,
  title={Resolvent-based approach for H 2-optimal estimation and control: an application to the cylinder flow},
  author={Jin, B. and Illingworth, S.J. and Sandberg, R.D.},
  journal={Theoretical and Computational Fluid Dynamics},
  volume={36},
  number={3},
  pages={491--515},
  year={2022},
  publisher={Springer}
}

@book{lehoucq1998arpack,
  title={ARPACK users' guide: solution of large-scale eigenvalue problems with implicitly restarted Arnoldi methods},
  author={Lehoucq, R.B. and Sorensen, D.C. and Yang, C.},
  year={1998},
  publisher={SIAM}
}

@article{davis2004algorithm,
  title={Algorithm 832: UMFPACK V4. 3---an unsymmetric-pattern multifrontal method},
  author={Davis, T.A.},
  journal={ACM Transactions on Mathematical Software (TOMS)},
  volume={30},
  number={2},
  pages={196--199},
  year={2004},
  publisher={ACM New York, NY, USA}
}

@book{chicone2006ordinary,
  title={Ordinary differential equations with applications},
  author={Chicone, C. C.},
  volume={34},
  year={2006},
  publisher={Springer}
}

@article{mizushima2005instability,
  title={Instability and transition of flow past two tandem circular cylinders},
  author={Mizushima, Jiro and Suehiro, Norihisa},
  journal={Physics of Fluids},
  volume={17},
  number={10},
  year={2005},
  publisher={AIP Publishing}
}

@article{wang2022first,
  title={First instability of the flow past two tandem cylinders with different diameters},
  author={Wang, J. and Shan, X. and Liu, J.},
  journal={Physics of Fluids},
  volume={34},
  number={7},
  year={2022},
  publisher={AIP Publishing}
}

@article{liu2024primary,
  title={Primary instability, sensitivity and active control of flow past two tandem circular cylinders},
  author={Liu, Z. and Zhou, L. and Tang, H. and Wang, Z. and Zhao, F. and Ji, X. and Zhang, H.},
  journal={Ocean Engineering},
  volume={294},
  pages={116863},
  year={2024},
  publisher={Elsevier}
}

@article{sharman2005numerical,
  title={Numerical predictions of low Reynolds number flows over two tandem circular cylinders},
  author={Sharman, B. and Lien, F.-S. and Davidson, L. and Norberg, C.},
  journal={International Journal for Numerical Methods in Fluids},
  volume={47},
  number={5},
  pages={423--447},
  year={2005},
  publisher={Wiley Online Library}
}

@article{latrobe2024flow,
  title={Flow control over tandem cylinders using plasma actuators},
  author={Latrobe, B. and Ohanu, E. G. and Fernandez, E. and Bhattacharya, S.},
  journal={Experimental Thermal and Fluid Science},
  volume={159},
  pages={111274},
  year={2024},
  publisher={Elsevier}
}

@article{xie2023applying,
  title={Applying reinforcement learning to mitigate wake-induced lift fluctuation of a wall-confined circular cylinder in tandem configuration},
  author={Xie, Z. and Hu, H. and Chen, J. and Song, J. and Lu, T. and Ren, F.},
  journal={Physics of Fluids},
  volume={35},
  number={5},
  year={2023},
  publisher={AIP Publishing}
}

@article{carmo2010possible,
  title={Possible states in the flow around two circular cylinders in tandem with separations in the vicinity of the drag inversion spacing},
  author={Carmo, B. S. and Meneghini, J. R. and Sherwin, S. J.},
  journal={Physics of Fluids},
  volume={22},
  number={5},
  year={2010},
  publisher={AIP Publishing}
}

@article{zdravkovich1985flow,
  title={Flow induced oscillations of two interfering circular cylinders},
  author={Zdravkovich, M.M.},
  journal={Journal of Sound and Vibration},
  volume={101},
  number={4},
  pages={511--521},
  year={1985},
  publisher={Elsevier}
}

@article{hetz1991vortex,
  title={Vortex shedding over five in-line cylinders cylinders},
  author={Hetz, A.A. and Dhaubhadel, M.N. and Telionis, D.P.},
  journal={Journal of fluids and structures},
  volume={5},
  number={3},
  pages={243--257},
  year={1991},
  publisher={Elsevier}
}

@book{kuznetsov1998elements,
  title={Elements of applied bifurcation theory},
  author={Kuznetsov, Y. A.},
  year={1998},
  publisher={Springer}
}

@article{wang2010secondary,
  title={Secondary vortex street in the wake of two tandem circular cylinders at low Reynolds number},
  author={Wang, Si-Ying and Tian, Fang-Bao and Jia, Lai-Bing and Lu, Xi-Yun and Yin, Xie-Zhen},
  journal={Physical Review E—Statistical, Nonlinear, and Soft Matter Physics},
  volume={81},
  number={3},
  pages={036305},
  year={2010},
  publisher={APS}
}

@article{symon2018non,
  title={Non-normality and classification of amplification mechanisms in stability and resolvent analysis},
  author={Symon, S.and Rosenberg, K. and Dawson, S.T.M. and McKeon, B.J.},
  journal={Physical Review Fluids},
  volume={3},
  number={5},
  pages={053902},
  year={2018},
  publisher={APS}
}

@article{jin2021energy,
  title={Energy transfer mechanisms and resolvent analysis in the cylinder wake},
  author={Jin, B. and Symon, S. and Illingworth, S. J.},
  journal={Physical Review Fluids},
  volume={6},
  number={2},
  pages={024702},
  year={2021},
  publisher={APS}
}

@article{ma2024suppression,
  title={Suppression of oscillatory fluid forces in cylinder wake: Optimal jet control position designed through resolvent analysis},
  author={Ma, R. and Gao, C. and Ren, K. and Yuan, H. and Zhang, W.},
  journal={Physics of Fluids},
  volume={36},
  number={7},
  year={2024},
  publisher={AIP Publishing}
}

@article{MR3043640,
  AUTHOR = {Hecht, F.},
  TITLE = {New development in FreeFem++},
  JOURNAL = {J. Numer. Math.},
  FJOURNAL = {Journal of Numerical Mathematics},
  VOLUME = {20}, YEAR = {2012},
  NUMBER = {3-4}, PAGES = {251--265},
  ISSN = {1570-2820},
  MRCLASS = {65Y15},
  MRNUMBER = {3043640},
  URL = {https://freefem.org/}
}
\end{document}